%% file: ir1.tex
\documentclass[a4paper,reqno]{amsart}
\usepackage[centertags]{amsmath}

\usepackage{amssymb}
\usepackage{amsthm}
\usepackage{latexsym}

\numberwithin{equation}{section}
 
\newcommand{\al}{\alpha}
\newcommand{\be}{\beta}
\newcommand{\ga}{\gamma}

\newcommand{\si}{\sigma}
\newcommand{\ro}{\rho}

\newcommand{\vp}{\varphi}
\newcommand{\vep}{\varepsilon}
\newcommand{\om}{\omega}
\newcommand{\ti}[1]{\tilde{#1}}
\newcommand{\wti}[1]{\widetilde{#1}}
\newcommand{\pal}{{^{\scriptstyle{,}}\!\alpha}}
\newcommand{\pbe}{{^{\scriptstyle{,}}\!\beta}}
\newcommand{\pga}{{^{\scriptstyle{,}}\!\gamma}}

\newcommand{\pxi}{{^{\scriptstyle{,}}\!\xi}}
\newcommand{\pro}{{^{\scriptstyle{,}}\!\rho}}

\newcommand{\pep}{{^{\scriptstyle{,}}\!\varepsilon}}

\newcommand{\psig}{{^{\scriptstyle{,}}\!\sigma}}
\newcommand{\pmu}{{^{\scriptstyle{,}}\!\mu}}
\newcommand{\pnu}{{^{\scriptstyle{,}}\!\nu}}

\newcommand{\tot}{\dagger}
\newcommand{\abs}[1]{\mathsf{#1}}
\newcommand{\contract}[1]{%
\overset{\makebox%
          {\, \rule[-.27em]{.3pt}{.3em}\hrulefill%
           \rule[-.27em]{.3pt}{.3em}\!\, }}%
        {\rule{0pt}{0.75em}#1}}%
\begin{document}

\title{Infrared Behaviour of Massive Scalar Matter coupled to Gravity}

\author[M. Wellmann] {{\large Mark Wellmann}\\
$\phantom{}$\\
$\phantom{}$\\
Institut f\"ur Theoretische Physik\\
der Universit\"at Z\"urich \\
Winterthurerstrasse 190  \\
CH-8057 Z\"urich Switzerland\\
$\phantom{}$\\
{\tt wellmann@physik.unizh.ch}}
%\date{\today}

\begin{abstract}
In the framework of causal perturbation theory we consider a massive scalar field coupled to gravity. In the field theoretic approach to quantum gravity (QG) we start with a massless second rank tensor field. This tensor field is then quantized in a covariant way in Minkowski space. This article deals with the adiabatic limit for graviton radiative corrections in a scattering process of two massive scalar particles. We compute the differential cross-section for bremsstrahlung processes in which one of the outgoing particles emits a graviton of low energy, a so called soft graviton. Since the emited graviton will not be detected we have to integrate over all soft gravitons.   
\end{abstract}

\maketitle
%#*#*#*#*#*#*#*#*#*#*#*#*#*#*#*#*#*#*#*#*#*#*#*#*#*#*#*#*#*#*#*#*#*#*#*#*#*#*#*#*#*#*#*#*#*#*#*#*#*#*#*#*#*#*#*#*#*#*#*#*#*#*#*#*#*#*#*#*#*#*#*#*#*#*#*#*#*#*#*#*#
\section{Introduction}
The study of the infrared problem in quantum gravity goes mainly back to the work of Weinberg \cite{wei:iphgrav} who has investigated the infrared behaviour of virtual and real soft graviton emission processes. The transition amplitude for a process in which an emission of a single graviton occurs is logarithmically divergent. In the sum over an infinite number of soft gravitons emitted he claims that infrared divergencies in the transition amplitude cancel to every order of perturbation theory. In his treatment he considers virtual and real bremsstrahlung seperately. For the \emph{virtual} ones he obtained in the sum over an  infinte number of bremsstrahlungs processes a transition amplitude which is proportional to some power of the lower cutoff parameter $\lambda$. So this rate will vanish in the limit $\lambda\rightarrow 0$. For an infinite number of \emph{real} processes he obtained a transition amplitude which is proportional to the same negative power in $\lambda$. So the cutoff disappears if virtual and real processes are combined. From our point of view we think that the problem can not be treated in this rather simple way for the following reasons. The structure of the problem is the same as in quantum electrodynamics (QED). Here it is well known that the infrared divergencies cancel in lowest order but in higher orders one has to deal with subdivergences in an arbitrary Feynman diagram. The question of wheather or not these divergencies cancel to every order of perturbation theory is a topic which had long been investigated, see \cite{yfs:idp,gy:itidp}, but is still under discussion \cite{stei:pqed}. As far as we know the cancellation of these divergencies is not at all clear. So we think that one has to investigate these processes carefully order by order.      

In this article we will reinvestigate the infrared problem in quantum gravity in the framework of causal perturbation theory. Especially we consider the emission of a real soft graviton in two particle scattering. We use the method of adiabatic switching with a testfunction from the Schwartz space which cuts off the interaction at large distances in spacetime. This makes all expressions well defined during the calculation and it turns out to be a natural infrared cutoff. We therefore avoid the introduction of a graviton mass. The latter, although widely used in the literature, is less satisfactory on physical grounds because it modifies the interaction at short distances, too. It is not at all clear that the different infrared regularizations give the same result for observable quantities. To study this question we want to calculate the differential cross section for bremsstrahlung from first principles.  

Causal perturbation theory goes back to ideas of Bogoliubov \cite{bog:itqf} and was carefully developed by Epstein and Glaser \cite{eg:rlp} in the seventies. Later Scharf \cite{sch:qed} applied it successfully to QED. The idea is to fix the first order interaction and then to construct higher orders of the scattering-matrix $S$ by induction using free fields only. Since the free quantum fields as basic objects are operator valued distributions the scattering-matrix will be constructed as an operator valued functional on some testfunction space. The most important ingredient for the inductive construction is the causality requirement, which roughly states that the scattering-matrix of a sum of two testfunctions factorizes if the supports of the testfunctions can be separated in time. As testfunction space one considers for convenience the Schwartz space of functions of rapid decrease, since this space is invariant under Fourier transformation. Then all expressions are well defined tempered distributions. In doing this we cut off the interaction at large but finite distances which is unphysical in most cases and has to be removed at the end. This is the so called adiabatic limit where we take the limit that the testfunction goes to a constant. In this way we investigate the long range behaviour of the theory.

%#*#*#*#*#*#*#*#*#*#*#*#*#*#*#*#*#*#*#*#*#*#*#*#*#*#*#*#*#*#*#*#*#*#*#*#*#*#*#*#*#*#*#*#*#*#*#*#*#*#*#*#*#*#*#*#*#*#*#*#*#*#*#*#*#*#*#*#*#*#*#*#*#*#*#*#*#*#*#*#*#
\section{Free fields}       
We consider a symmetric second rank tensor field $h^{\mu\nu}(x)$, which is a solution to the wave equation, i.e. $\Box h^{\mu\nu}(x)=0$. All tensor indices are raised and lowered with the Minkowski metric $\eta_{\mu\nu}$ with components $\eta_{00}=+1=-\eta_{ii},\ i=1,2,3$. We denote by $kx\equiv k\cdot x=\eta_{\mu\nu}k^{\mu}x^{\nu}$ the Minkowski scalar product of two four vectors. Throughout the paper equal indizes in one expression are always be properly contracted with the Minkowski metric irrespective of their position. We quantize the graviton field $h^{\mu\nu}(x)$ by imposing the Lorentz covariant commutation relations
\begin{equation}
  \bigl[h^{\al\be}(x),h^{\mu\nu}(y)\bigr]=-ib^{\al\be\mu\nu}D_0(x-y)
\label{hh}
\end{equation}
where the tensor $b^{\al\be\mu\nu}$ is defined by
\begin{equation}
  b^{\al\be\mu\nu}=\frac{1}{2}\bigl(\eta^{\al\mu}\eta^{\be\nu}+\eta^{\al\nu}\eta^{\be\mu}-\eta^{\al\be}\eta^{\mu\nu}\bigr)
\label{b-tensor}
\end{equation}
and $D_0(x-y)$ is the massless Pauli-Jordan distribution. We can write down the Fourier-representation of the field $h^{\mu\nu}$. It is given by
\begin{equation}
  h^{\al\be}(x)=(2\pi)^{-3/2}\int \frac{d^3\vec{k}}{\sqrt{2\omega(\vec{k})}}\bigl(a^{\al\be}(\vec{k})\exp(-ikx)+a^{\al\be}(\vec{k})^{\tot}\exp(+ikx)\bigr)
\label{h-field}
\end{equation}
Here $\omega(\vec{k})=|\vec{k}|$ and $a^{\al\be}(\vec{k})$, $a^{\al\be}(\vec{k})^{\tot}$ are annihilation and creation operators on a bosonic Fock-space. From (\ref{hh}) we find that they have the following commutation relations
\begin{equation}
  \bigl[a^{\al\be}(\vec{k}),a^{\mu\nu}(\vec{k}^{\prime})^{\tot}\bigr]=b^{\al\be\mu\nu}\delta^{(3)}(\vec{k}-\vec{k}^{\prime})
\label{h-com}
\end{equation}
It should be emphasized that this quantization procedure is only formal in the sense that we don't describe physical gravitons, since our $h^{\mu\nu}$  contains too many degrees of freedom for a real gravitational field. There must be imposed further conditions like a specific gauge condition, etc. in order to obtain physical gravitons. This was done in \cite{gr:cqg2}.    
The massive scalar field $\vp(x)$ obeys the Klein-Gordan equation $(\Box +m^2)\vp(x)=0$. It has the following representation in terms of annihilation- and creation-operators:
\begin{equation}
  \vp(x)=(2\pi)^{-3/2}\int \frac{d^3\vec{p}}{\sqrt{2\omega(\vec{p})}}\bigl(a(\vec{p})\exp(-ipx)+\ti{a}(\vec{p})^{\tot}\exp(+ipx)\bigr)
\label{phi-field}
\end{equation}
 The energy $\om(\vec{p})$ is as usual given by $\omega(\vec{p})=\sqrt{\vec{p_{}}^2+m^2}$. The operator $a(\vec{p})$ annihilates a particle with momentum $\vec{p}$ whereas $\ti{a}(\vec{p})^{\tot}$ creates an antiparticle with momentum $\vec{p}$. These operators obey the following commutation relations
\begin{equation}
  \begin{split}
    \bigl[a(\vec{p}),{a(\vec{p_{}}^{\prime})}^{\tot}\bigr] & =\delta^{(3)}(\vec{p}-\vec{p_{}}^{\prime}) \\
    \bigl[\ti{a}(\vec{p}),{\ti{a}(\vec{p_{}}^{\prime})}^{\tot}\bigr] & = \delta^{(3)}(\vec{p}-\vec{p_{}}^{\prime}) \\ 
  \end{split}
\label{phi-com}
\end{equation}
and all the other commutators vanish.

%#*#*#*#*#*#*#*#*#*#*#*#*#*#*#*#*#*#*#*#*#*#*#*#*#*#*#*#*#*#*#*#*#*#*#*#*#*#*#*#*#*#*#*#*#*#*#*#*#*#*#*#*#*#*#*#*#*#*#*#*#*#*#*#*#*#*#*#*#*#*#*#*#*#*#*#*#*#*#*#*#*#
\section{Interaction of massive scalar matter with gravitation}
The energy-momentum tensor of $\vp(x)$  is given by
\begin{equation}
  T^{\mu\nu}_m(x)=\ :\Bigl(\vp(x)^{\tot}_{\pmu}\vp(x)_{\pnu}+\vp(x)^{\tot}_{\pnu}\vp(x)_{\pmu}-\eta_{\mu\nu}\bigl(\vp(x)^{\tot}_{\pxi}\vp(x)_{\pxi}-m^2\vp(x)^{\tot}
                  \vp(x)\bigr)\Bigr):
\end{equation}
The interaction with the gravitational field will be described through the following first order coupling \cite{gr:scmqg}
\begin{equation}
  T_1(x)=\frac{i}{2}\kappa :h^{\al\be}(x)b_{\al\be\mu\nu}T^{\mu\nu}_m(x):
\end{equation}
where $\kappa=\sqrt{32\pi G}$ with Newton's constant $G$. This leads to the following explicit form of the first order interaction
\begin{equation}
  T_1(x)=\frac{i}{2}\kappa\Bigl[2:h^{\al\be}(x)\vp(x)^{\tot}_{\pal}\vp(x)_{\pbe}:-m^2:h(x)\vp(x)^{\tot}\vp(x):\Bigr]
  \label{T1}
\end{equation}
where we have introduced the trace of the graviton field $h(x)=h(x)^{\al}_{\al}$.

%#*#*#*#*#*#*#*#*#*#*#*#*#*#*#*#*#*#*#*#*#*#*#*#*#*#*#*#*#*#*#*#*#*#*#*#*#*#*#*#*#*#*#*#*#*#*#*#*#*#*#*#*#*#*#*#*#*#*#*#*#*#*#*#*#*#*#*#*#*#*#*#*#*#*#*#*#*#*#*#*#*#*#
\section{$S$-matrix to third order}
For the calculation of the bremsstrahlung process we need to know the explicit form of the $S$-matrix up to third order in the coupling constant $\kappa$. This means that we have to construct the time-ordered product $T_3(x_1,x_2,x_3)$. According to the inductive construction of Epstein and Glaser we proceed by first calculating $T_2(x_1,x_2)$ which will be done in the first subsection and after that we calculate the relevant terms, i.e. terms with four external scalar field operators and one graviton field operator, in the next subsection.

%#*#*#*#*#*#*#*#*#*#*#*#*#*#*#*#*#*#*#*#*#*#*#*#*#*#*#*#*#*#*#*#*#*#*#*#*#*#*#*#*#*#*#*#*#*#*#*#*#*#*#*#*#*#*#*#*#*#*#*#*#*#*#*#*#*#*#*#*#*#*#*#*#*#*#*#*#*#*#*#*#*#*#*#
\subsection{Time-ordered product to second order}
For the construction of the time-ordered product $T_2(x_1,x_2)$ we start with the first order term(\ref{T1}) and apply the inductive construction due to Epstein-Glaser as in Scharf's book \cite{sch:qed}. First of all we have to build the two distributions
\begin{equation}
  A_2^{\prime}(x_1,x_2)=-T_1(x_1)T_1(x_2)
\end{equation}
and 
\begin{equation}
  R_2^{\prime}(x_1,x_2)=-T_1(x_2)T_1(x_1)
\end{equation}
Since $A_2^{\prime}$ and $R_2^{\prime}$ are products of normally ordered field operators we apply Wick's theorem to obtain a normally ordered expression. We are interested in scattering processes so we have to concentrate on terms with zero or one contraction only. The contraction of the free field operators are defined by
\begin{equation}
  \begin{split}
    \contract{h^{\al\be}(x_1)h^{\mu\nu}}(x_2) := & -ib^{\al\be\mu\nu}D^{(+)}_0(x_1-x_2)\\
    \contract{\vp(x_1)\vp}(x_2) := & -iD^{(+)}_m(x_1-x_2)\\
  \end{split}
\end{equation}
where $D^{(+)}_m(x)$ is the positive frequency part of the massive Pauli-Jordan distribution. For the splitting operation we need the difference 
\begin{equation}
  D_2(x_1,x_2)=R_2^{\prime}(x_1,x_2)-A_2^{\prime}(x_1,x_2)
\end{equation}

This distribution has causal support and it is given by
\begin{equation}
  \begin{split}
    D_2(x_1,x_2) = & \ i\frac{\kappa^2}{4}\Bigl[4:h^{\al\be}(x_2)h^{\ro\si}(x_1)\vp(x_2)_{\pbe}\vp(x_1)^{\tot}_{\pro}:\partial^{x_2}_{\al}\partial^{x_1}_{\si}D_m(x_1-x_2) \\
                   & +4:h^{\al\be}(x_2)h^{\ro\si}(x_1)\vp(x_2)^{\tot}_{\pal}\vp(x_1)_{\psig}:\partial^{x_2}_{\be}\partial^{x_1}_{\ro}D_m(x_1-x_2) \\
                   & -2m^2:h^{\al\be}(x_2)h(x_1)\vp(x_2)_{\pbe}\vp(x_1)^{\tot}:\partial^{x_2}_{\al}D_m(x_1-x_2) \\
                   & -2m^2:h^{\al\be}(x_2)h(x_1)\vp(x_2)^{\tot}_{\pal}\vp(x_1):\partial^{x_2}_{\be}D_m(x_1-x_2) \\
                   & -2m^2:h(x_2)h^{\ro\si}(x_1)\vp(x_2)\vp(x_1)^{\tot}_{\pro}:\partial^{x_1}_{\si}D_m(x_1-x_2) \\
                   & -2m^2:h(x_2)h^{\ro\si}(x_1)\vp(x_2)^{\tot}\vp(x_1)_{\psig}:\partial^{x_1}_{\ro}D_m(x_1-x_2) \\
                   & +m^4:h(x_2)h(x_1)\vp(x_2)\vp(x_1)^{\tot}:D_m(x_1-x_2) \\
                   & +m^4:h(x_2)h(x_1)\vp(x_2)^{\tot}\vp(x_1):D_m(x_1-x_2) \\
                   & + 4:\vp(x_2)^{\tot}_{\pro}\vp(x_2)_{\psig}\vp(x_1)^{\tot}_{\pal}\vp(x_1)_{\pbe}:b^{\al\be\ro\si}D_0(x_1-x_2) \\
                   & +2m^2:\vp(x_2)^{\tot}_{\pro}\vp(x_2)_{\pro}\vp(x_1)^{\tot}\vp(x_1):D_0(x_1-x_2) \\
                   & +2m^2:\vp(x_2)^{\tot}\vp(x_2)\vp(x_1)^{\tot}_{\pal}\vp(x_1)_{\pal}:D_0(x_1-x_2) \\
                   & -4m^4:\vp(x_2)^{\tot}\vp(x_2)\vp(x_1)^{\tot}\vp(x_1):D_0(x_1-x_2)\Bigr] \\
  \end{split}
\end{equation}
To obtain the time-ordered product $T_2$ we have to split the numerical distribution in $D_2$ according to it's singular order (see the Appendix for the definition of the singular order). The singular order of the Pauli-Jordan distribution is $\omega=-2$, so the splitting is trivial and we can use the formula
\begin{equation}
  D_m(x)=D^{ret}_m(x)-D^{av}_m(x).
\end{equation}
The retarded distribution $R_2(x_1,x_2)$ is then given by $D_2(x_1,x_2)$ if we replace all Pauli-Jordan distributions by the retarded ones $D^{ret}$. Finally the time-ordered distribution $T_2(x_1,x_2)$ is given by
\begin{equation}
  T_2(x_1,x_2)=R_2(x_1,x_2)-R_2^{\prime}(x_1,x_2)
\end{equation}
For later reference we give here the explicit expression
\begin{equation}
  \begin{split}
     T_2(x_1,x_2) = & \ i\frac{\kappa^2}{4}\Bigl[4:h^{\al\be}(x_2)h^{\ro\si}(x_1)\vp(x_2)_{\pbe}\vp(x_1)^{\tot}_{\pro}:\partial^{x_2}_{\al}\partial^{x_1}_{\si}D^F_m(x_1-x_2) \\
                   & +4:h^{\al\be}(x_2)h^{\ro\si}(x_1)\vp(x_2)^{\tot}_{\pal}\vp(x_1)_{\psig}:\partial^{x_2}_{\be}\partial^{x_1}_{\ro}D^F_m(x_1-x_2) \\
                   & -2m^2:h^{\al\be}(x_2)h(x_1)\vp(x_2)_{\pbe}\vp(x_1)^{\tot}:\partial^{x_2}_{\al}D^F_m(x_1-x_2) \\
                   & -2m^2:h^{\al\be}(x_2)h(x_1)\vp(x_2)^{\tot}_{\pal}\vp(x_1):\partial^{x_2}_{\be}D^F_m(x_1-x_2) \\
                   & -2m^2:h(x_2)h^{\ro\si}(x_1)\vp(x_2)\vp(x_1)^{\tot}_{\pro}:\partial^{x_1}_{\si}D^F_m(x_1-x_2) \\
                   & -2m^2:h(x_2)h^{\ro\si}(x_1)\vp(x_2)^{\tot}\vp(x_1)_{\psig}:\partial^{x_1}_{\ro}D^F_m(x_1-x_2) \\
                   & +m^4:h(x_2)h(x_1)\vp(x_2)\vp(x_1)^{\tot}:D^F_m(x_1-x_2) \\
                   & +m^4:h(x_2)h(x_1)\vp(x_2)^{\tot}\vp(x_1):D^F_m(x_1-x_2) \\
                   & + 4:\vp(x_2)^{\tot}_{\pro}\vp(x_2)_{\psig}\vp(x_1)^{\tot}_{\pal}\vp(x_1)_{\pbe}:b^{\al\be\ro\si}D^F_0(x_1-x_2) \\
                   & +2m^2:\vp(x_2)^{\tot}_{\pro}\vp(x_2)_{\pro}\vp(x_1)^{\tot}\vp(x_1):D^F_0(x_1-x_2) \\
                   & +2m^2:\vp(x_2)^{\tot}\vp(x_2)\vp(x_1)^{\tot}_{\pal}\vp(x_1)_{\pal}:D^F_0(x_1-x_2) \\
                   & -4m^4:\vp(x_2)^{\tot}\vp(x_2)\vp(x_1)^{\tot}\vp(x_1):D^F_0(x_1-x_2) \\
                   & -4:h^{\al\be}(x_1)\vp(x_1)^{\tot}_{\pal}\vp(x_1)_{\pbe}h^{\ro\si}(x_2)\vp(x_2)^{\tot}_{\pro}\vp(x_2)_{\psig}: \\
                   & +2m^2:h^{\al\be}(x_1)\vp(x_1)^{\tot}_{\pal}\vp(x_1)_{\pbe}h(x_2)\vp(x_2)^{\tot}\vp(x_2): \\
                   & +2m^2:h(x_1)\vp(x_1)^{\tot}\vp(x_1)h^{\ro\si}(x_2)\vp(x_2)^{\tot}_{\pro}\vp(x_2)_{\psig}: \\
                   & -m^4:h(x_1)\vp(x_1)^{\tot}\vp(x_1)h(x_2)\vp(x_2)^{\tot}\vp(x_2):\Bigr] \\
  \end{split}
\label{T2}
\end{equation}
This result will be needed in the next subsection for the calculation of $T_3$.

%#*#*#*#*#*#*#*#*#*#*#*#*#*#*#*#*#*#*#*#*#*#*#*#*#*#*#*#*#*#*#*#*#*#*#*#*#*#*#*#*#*#*#*#*#*#*#*#*#*#*#*#*#*#*#*#*#*#*#*#*#*#*#*#*#*#*#*#*#*#*#*#*#*#*#*#*#*#*#
\subsection{Time-ordered product to third order}
In the previous subsection we have calculated the distribution $T_2$(\ref{T2}) from the given first order coupling $T_1$(\ref{T1}). Now to obtain the time-ordered product to third order we proceed in much the same way as in the calculation of $T_2$. We calculate the distributions $R_3^{\prime}(x_1,x_2,x_3)$ and $A_3^{\prime}(x_1,x_2,x_3)$ which are given by
\begin{equation}
  R_3^{\prime}(x_1,x_2,x_3)=T_2(x_1,x_3)\wti{T}_1(x_2)+T_2(x_2,x_3)\wti{T}_1(x_1)+T_1(x_3)\wti{T}_2(x_1,x_2)
\end{equation}
and
\begin{equation}
  A_3^{\prime}(x_1,x_2,x_3)=\wti{T}_1(x_2)T_2(x_1,x_3)+\wti{T}_1(x_1)T_2(x_2,x_3)+\wti{T}_2(x_1,x_2)T_1(x_3)
\end{equation}
where $\wti{T}_1$ and $\wti{T}_2$ are defined by
\begin{equation}
  \wti{T}_1(x_i):=-T_1(x_i),\quad i=1,2,3
\end{equation}
and
\begin{equation}
 \wti{T}_2(x_i,x_j):=-T_2(x_i,x_j)+T_1(x_i)T_1(x_j)+T_1(x_j)T_1(x_i),\quad i,j=1,2,3
\end{equation}
It should be noticed that we only collect those terms which have two contractions involving all the spacetime points $x_1,x_2$ and $x_3$, i.e. there is no product of contraction functions with the same argument. After we have built the causal difference $D_3=R_3^{\prime}-A_3^{\prime}$ we split the numerical part into $R_3$ with support in the backward light cone and $A_3$ with support in the forward light cone. The time-ordered distribution $T_3$ is then given by
\begin{equation}
  T_3(x_1,x_2,x_3)=R_3(x_1,x_2,x_3)-R_3^{\prime}(x_1,x_2,x_3)
\end{equation}
Because the calculation is staightforward and not very enlightning we give only the result
\begin{equation}
  \begin{split}
    T_3(x_1,x_2,x_3) = & \ \frac{i\kappa^3}{8}\sum_{\pi\in S_3}\Biggl[8\Bigl[:h^{\ro\si}(x_{\pi(1)})\vp(x_{\pi(1)})^{\tot}_{\psig}\vp(x_{\pi(2)})^{\tot}_{\pga}
                         \vp(x_{\pi(2)})_{\pep}\vp(x_{\pi(3)})_{\pal}:\\
                       & \ +:h^{\ro\si}(x_{\pi(1)})\vp(x_{\pi(1)})_{\psig}\vp(x_{\pi(2)})^{\tot}_{\pga}\vp(x_{\pi(2)})_{\pep}\vp(x_{\pi(3)})^{\tot}_{\pal}:\Bigr]\\ 
                       & \ \times b^{\al\be\ga\vep}D^F_0(x_{\pi(2)}-x_{\pi(3)})\partial^{x_{\pi(3)}}_{\be}\partial^{x_{\pi(1)}}_{\ro}D^F_m(x_{\pi(1)}-x_{\pi(3)})\\
                       & \ +4m^2\Bigl[:h^{\ro\si}(x_{\pi(1)})\vp(x_{\pi(1)})^{\tot}_{\pro}\vp(x_{\pi(2)})^{\tot}\vp(x_{\pi(2)})\vp(x_{\pi(3)})_{\pal}:\\
                       & \ +:h^{\ro\si}(x_{\pi(1)})\vp(x_{\pi(1)})_{\pro}\vp(x_{\pi(2)})^{\tot}\vp(x_{\pi(2)})\vp(x_{\pi(3)})^{\tot}_{\pal}:\Bigr]\\
                       & \ \times D^F_0(x_{\pi(2)}-x_{\pi(3)})\partial^{x_{\pi(3)}}_{\al}\partial^{x_{\pi(1)}}_{\si}D^F_m(x_{\pi(1)}-x_{\pi(3)})\\
                       & \ +4m^2\Bigl[:h^{\ro\si}(x_{\pi(1)})\vp(x_{\pi(1)})^{\tot}_{\pro}\vp(x_{\pi(2)})^{\tot}_{\pep}\vp(x_{\pi(2)})_{\pep}\vp(x_{\pi(3)}):\\
                       & \ +:h^{\ro\si}(x_{\pi(1)})\vp(x_{\pi(1)})_{\pro}\vp(x_{\pi(2)})^{\tot}_{\pep}\vp(x_{\pi(2)})_{\pep}\vp(x_{\pi(3)})^{\tot}:\Bigr]\\
                       & \ \times D^F_0(x_{\pi(3)}-x_{\pi(2)})\partial^{x_{\pi(1)}}_{\si}D^F_m(x_{\pi(1)}-x_{\pi(3)})\\
                       & \ +4m^2\Bigl[:h(x_{\pi(1)})\vp(x_{\pi(1)})^{\tot}\vp(x_{\pi(2)})^{\tot}_{\pga}\vp(x_{\pi(2)})_{\pep}\vp(x_{\pi(3)})_{\pbe}:\\
                       & \ +:h(x_{\pi(1)})\vp(x_{\pi(1)})\vp(x_{\pi(2)})^{\tot}_{\pga}\vp(x_{\pi(2)})_{\pep}\vp(x_{\pi(3)})^{\tot}_{\pbe}:\Bigr]\\
                       & \ \times b^{\al\be\ga\vep}D^F_0(x_{\pi(3)}-x_{\pi(2)})\partial^{x_{\pi(3)}}_{\al}D^F_m(x_{\pi(1)}-x_{\pi(3)})\\
                       & \ +2m^4\Bigl[:h(x_{\pi(1)})\vp(x_{\pi(1)})^{\tot}\vp(x_{\pi(2)})^{\tot}_{\pga}\vp(x_{\pi(2)})_{\pga}\vp(x_{\pi(3)}):\\
                       & \ +:h(x_{\pi(1)})\vp(x_{\pi(1)})\vp(x_{\pi(2)})^{\tot}_{\pga}\vp(x_{\pi(2)})_{\pga}\vp(x_{\pi(3)})^{\tot}:\Bigr]\\
                       & \ \times D^F_0(x_{\pi(3)}-x_{\pi(2)})D^F_m(x_{\pi(1)}-x_{\pi(3)})\\
                       & \ +2m^4\Bigl[:h(x_{\pi(1)})\vp(x_{\pi(1)})^{\tot}\vp(x_{\pi(2)})^{\tot}\vp(x_{\pi(2)})\vp(x_{\pi(3)})_{\pbe}:\\
                       & \ +:h(x_{\pi(1)})\vp(x_{\pi(1)})\vp(x_{\pi(2)})^{\tot}\vp(x_{\pi(2)})\vp(x_{\pi(3)})^{\tot}_{\pbe}:\Bigr]\\
                       & \ \times D^F_0(x_{\pi(3)}-x_{\pi(2)})\partial^{x_{\pi(3)}}_{\be}D^F_m(x_{\pi(1)}-x_{\pi(3)})\\
                       & \ +8m^4\Bigl[:h^{\ro\si}(x_{\pi(1)})\vp(x_{\pi(1)})^{\tot}_{\pro}\vp(x_{\pi(2)})^{\tot}\vp(x_{\pi(2)})\vp(x_{\pi(3)}):\\
                       & \ +:h^{\ro\si}(x_{\pi(1)})\vp(x_{\pi(1)})_{\pro}\vp(x_{\pi(2)})^{\tot}\vp(x_{\pi(2)})\vp(x_{\pi(3)})^{\tot}:\Bigr]\\
                       & \ \times D^F_0(x_{\pi(3)}-x_{\pi(2)})\partial^{x_{\pi(1)}}_{\si}D^F_m(x_{\pi(1)}-x_{\pi(3)})\\
                       & \ +4m^6\Bigl[:h(x_{\pi(1)})\vp(x_{\pi(1)})^{\tot}\vp(x_{\pi(2)})^{\tot}\vp(x_{\pi(2)})\vp(x_{\pi(3)}):\\
                       & \ +:h(x_{\pi(1)})\vp(x_{\pi(1)})\vp(x_{\pi(2)})^{\tot}\vp(x_{\pi(2)})\vp(x_{\pi(3)})^{\tot}:\Bigr]\\
                       & \ \times D^F_0(x_{\pi(3)}-x_{\pi(2)})D^F_m(x_{\pi(1)}-x_{\pi(3)})\Biggr]\\
  \end{split}
\label{T3}
\end{equation}
Due to the sum over all permutations of the indices this expression for $T_3$ is obviously symmetric in it's arguments as it is required by the definition of the time-ordered product. The third order $S$-matrix for the bremsstrahlungs processes is then given by
\begin{equation}
  S_3(g)=\int d^4x_1\ldots d^4x_3T_3(x_1,x_2,x_3)g(x_1)g(x_2)g(x_3),\quad g\in\mathcal{S}(\mathbb{R}^4)
\label{S-matrix}
\end{equation}

%#*#*#*#*#*#*#*#*#*#*#*#*#*#*#*#*#*#*#*#*#*#*#*#*#*#*#*#*#*#*#*#*#*#*#*#*#*#*#*#*#*#*#*#*#*#*#*#*#*#*#*#*#*#*#*#*#*#*#*#*#*#*#*#*#*#*#*#*#*#*#*#*#*#*#*#*#*#*#
\section{Two particle scattering with Bremsstrahlung}
\subsection{Towards the differential cross section--Step 1: $S$-matrix element}
In the preceeding section we have shown explicitly the construction of the time-ordered product to second and to third order. The $S$-matrix to third order was given by (\ref{S-matrix}). Now we consider the scattering process of two massive scalar particles in which a bremsstrahlungs graviton is emited in the final state. That is we want to calculate the expectation value of $S_3(g)$ between the following initial and final states
\begin{align}
   \Phi_i & = \int d^3\vec{p}_1d^3\vec{q}_1\hat{\psi}_1(\vec{p}_1)\hat{\psi}_2(\vec{q}_1)a(\vec{p}_1)^{\tot}a(\vec{q}_1)^{\tot}\Omega \\
   \Phi_f & = \int d^3\vec{k} d^3\vec{p}_2d^3\vec{q}_2\hat{\Psi}(\vec{p}_2,\vec{q}_2)\hat{\phi}^{\mu\nu}(\vec{k})a^{\mu\nu}(\vec{k})^{\tot}a(\vec{p}_2)^{\tot}
              a(\vec{q}_2)^{\tot}\Omega
\label{States}
\end{align}
These are wave packets in momentum space where $\hat{\psi}_i,\ i=1,2$ are one particle wave functions from $L^2(\mathbb{R}^3)$, $\hat{\Psi}$ is a two particle wave function from $L^2(\mathbb{R}^3\otimes\mathbb{R}^3)$ and $\hat{\phi}^{\mu\nu}$ is a tensor-valued square integrable wave function on $\mathbb{R}^3$. The vector $\Omega$ is the vacuum vector in the asymptotic Fock-Hilbert space. We want to calculate the following $S$-matrix element: 
\begin{equation}
  S_{fi}=(\Phi_f,S_3(g)\Phi_i)
\label{Sfi1}
\end{equation}
To show the details of the calculation we restrict ourself to the following term from the above computed $T_3$ (\ref{T3}) 
\begin{equation}
  \begin{split}
    T^{(1)}_3(x_1,x_2,x_3) := & \ i\kappa^3:h^{\ro\si}(x_1)\vp(x_3)_{\pbe}\vp(x_1)^{\tot}_{\pro}\vp(x_2)^{\tot}_{\pga}\vp(x_2)_{\pep}:\\
                              & \ b^{\al\be\ga\vep}D^F_0(x_2-x_3)\partial^{x_3}_{\al}\partial^{x_1}_{\si}D^F_m(x_1-x_3)\\
  \end{split}
\label{T3Term1}
\end{equation}
where the $(1)$ refers to this particular term of $T_3$. The $S$-matrix element corresponding to this contribution is then given by
\begin{equation}
  \begin{split}
    S_{fi}^{(1)} = & \int d^3\vec{p}_1d^3\vec{q_1}d^3\vec{k}d^3\vec{p}_2d^3\vec{q}_2d^4x_1\ldots d^4x_3\hat{\psi}_1(\vec{p}_1)\hat{\psi}_2(\vec{q}_1)
                     \hat{\Psi}(\vec{p}_2,\vec{q}_2)\hat{\phi}^{\mu\nu}(\vec{k}) \\
                   & \times g(x_1)\ldots g(x_3)\bigl(a^{\mu\nu}(\vec{k})^{\tot}a(\vec{p}_2)^{\tot}a(\vec{q}_2)^{\tot}\Omega , T^{(1)}_3(x_1,x_2,x_3)
                     a(\vec{p}_1)^{\tot}a(\vec{q}_1)^{\tot}\Omega\bigr)\\
  \end{split}
\label{Sfi2}
\end{equation}
Then we have to compute the vacuum expectation value
\begin{equation}
  \bigl(\Omega ,a(\vec{q_2})a(\vec{p_2})a^{\mu\nu}(\vec{k}):h^{\ro\si}(x_1)\vp(x_3)_{\pbe}\vp(x_1)^{\tot}_{\pro}\vp(x_2)^{\tot}_{\pga}\vp(x_2)_{\pep}:a(\vec{p_1})^{\tot}
  a(\vec{q_1})^{\tot}\Omega\bigr)
\label{vep}
\end{equation}
This can easily be evaluated by inserting the Fourier representations of the free field operators (\ref{h-field}) and (\ref{phi-field}) and the use of the commutation relations in momentum space (\ref{h-com}),(\ref{phi-com}). The result is given by
\begin{equation}
  \begin{split}
    \bigl(\Omega ,\ldots\Omega\bigr) = & \ (2\pi)^{-15/2}b^{\mu\nu\ro\si}\bigl(32\omega(\vec{k})\omega(\vec{p}_1)\omega(\vec{p}_2)\omega(\vec{q}_1)\omega(\vec{q}_2)
                                         \bigr)^{-1/2}\\
                                       & \Bigl[\exp\bigl[+i(kx_1-p_1x_3+q_2x_1+p_2x_2-q_1x_2)\bigr]\,p_1^{\be}\,q_2^{\ro}\,p_2^{\ga}\,q_1^{\vep} \\
                                       & +\exp\bigl[+i(kx_1-q_1x_3+q_2x_1+p_2x_2-p_1x_2)\bigr]\,q_1^{\be}\,q_2^{\ro}\,p_2^{\ga}\,p_1^{\vep}\\
                                       & +\exp\bigl[+i(kx_1-p_1x_3+p_2x_1+q_2x_2-q_1x_2)\bigr]\,p_1^{\be}\,p_2^{\ro}\,q_2^{\ga}\,q_1^{\vep}\\
                                       & +\exp\bigl[+i(kx_1-q_1x_3+p_2x_1+q_2x_2-p_1x_2)\bigr]\,q_1^{\be}\,p_2^{\ro}\,q_2^{\ga}\,p_1^{\vep}\Bigr]\\
  \end{split}
\end{equation}
The $S$-matrix element then reads
\begin{equation}
  \begin{split}
    S_{fi}^{(1)} = & \ i(2\pi)^{-15/2}\kappa^3\int d^3\vec{p}_1d^3\vec{q_1}d^3\vec{k}d^3\vec{p}_2d^3\vec{q}_2d^4x_1\ldots d^4x_3\hat{\psi}_1(\vec{p}_1)\hat{\psi}_2(\vec{q}_1)
                     \hat{\Psi}(\vec{p}_2,\vec{q}_2) \\
                   & \ \hat{\phi}^{\mu\nu}(\vec{k})g(x_1)\ldots g(x_3)b^{\al\be\ga\vep}D^F_0(x_2-x_3)\partial^{x_3}_{\al}\partial^{x_1}_{\si}D^F_m(x_1-x_3)b^{\mu\nu\ro\si}\\
                   & \ \bigl(32\omega(\vec{k})\omega(\vec{p}_1)\omega(\vec{p}_2)\omega(\vec{q}_1)\omega(\vec{q_2})\bigr)^{-1/2}\\
                   & \Bigl[\exp\bigl[+i(kx_1-p_1x_3+q_2x_1+p_2x_2-q_1x_2)\bigr]\,p_1^{\be}\,q_2^{\ro}\,p_2^{\ga}\,q_1^{\vep} \\
                   & +\exp\bigl[+i(kx_1-q_1x_3+q_2x_1+p_2x_2-p_1x_2)\bigr]\,q_1^{\be}\,q_2^{\ro}\,p_2^{\ga}\,p_1^{\vep}\\
                   & +\exp\bigl[+i(kx_1-p_1x_3+p_2x_1+q_2x_2-q_1x_2)\bigr]\,p_1^{\be}\,p_2^{\ro}\,q_2^{\ga}\,q_1^{\vep}\\
                   & +\exp\bigl[+i(kx_1-q_1x_3+p_2x_1+q_2x_2-p_1x_2)\bigr]\,q_1^{\be}\,p_2^{\ro}\,q_2^{\ga}\,p_1^{\vep}\Bigr]\\
  \end{split}
\label{Sfi3}
\end{equation} 
We introduce the abbreviations $I_1\ldots I_4$ for the four parts of $S_{fi}^{(1)}$. Then we consider the spatial integrations in $I_1$. They can be carried out by inserting the Fourier transforms of the testfunctions as well as the Feynman propagators:
\begin{equation}
  \begin{split}
    I_1 = & \int d^4x_1\ldots d^4x_3g(x_1)\ldots g(x_3)D^F_0(x_2-x_3)\partial^{x_3}_{\al}\partial^{x_1}_{\si}D^F_m(x_1-x_3)\\
          & \ \exp\bigl[+i(kx_1-p_1x_3+q_2x_1+p_2x_2-q_1x_2)\bigr]\\
        = & \ (2\pi)^{-10}\int d^4x_1\ldots d^4x_3d^4l_1\ldots d^4l_5\hat{g}(l_1)\ldots\hat{g}(l_3)\exp\bigl[-i(l_1x_1+l_2x_2+l_3x_3)\bigr] \\
          & \ \hat{D}^F_0(l_4)\exp\bigl[-il_4(x_2-x_3)\bigr]\partial^{x_3}_{\al}\partial^{x_1}_{\si}\hat{D}^F_m(l_5)\exp\bigl[-il_5(x_1-x_3)\bigr] \\
          & \ \exp\bigl[+i(kx_1-p_1x_3+q_2x_1+p_2x_2-q_1x_2)\bigr]\\
        = & \ (2\pi)^{-10}\int d^4x_1\ldots d^4x_3d^4l_1\ldots d^4l_5\hat{g}(l_1)\ldots\hat{g}(l_3)\exp\bigl[-i(l_1x_1+l_2x_2+l_3x_3)\bigr] \\
          & \ \exp\bigl[-il_4(x_2-x_3)\bigr]l_5^{\al}l_5^{\si}\hat{D}^F_m(l_5)\exp\bigl[-il_5(x_1-x_3)\bigr] \\
          & \ \exp\bigl[+i(kx_1-p_1x_3+q_2x_1+p_2x_2-q_1x_2)\bigr]\\
        = & \ (2\pi)^2\int d^4l_1\ldots d^4l_5\hat{g}(l_1)\ldots\hat{g}(l_3)\hat{D}^F_0(l_4)l_5^{\al}l_5^{\si}\hat{D}^F_m(l_5)\delta^{(4)}(l_1+l_5-k-q_2) \\
          & \ \delta^{(4)}(l_2+l_4-p_2+q_1)\delta^{(4)}(l_3-l_4-l_5+p_1)\\
  \end{split}
\label{I1}
\end{equation}
We can do the integrations w.r.t. $l_4$ and $l_5$ immediately and get
\begin{equation}
  \begin{split}
    I_1 = & \ (2\pi)^2\int d^4l_1\ldots d^4l_3\hat{g}(l_1)\ldots\hat{g}(l_3)\hat{D}^F_0(l_1+l_3+p_1-k-q_2)\\
          & \ [l_1^{\al}-p_1^{\al}+k^{\al}+q_2^{\al}][-l_1^{\si}-p_1^{\si}+k^{\si}+q_2^{\si}]\hat{D}^F_m(-l_1+k+q_2)\\
          & \ \delta^{(4)}(l_1+l_2+l_3+p_1+q_1-q_2-p_2-k)\\
  \end{split}
\end{equation}
Then the first term of $S^{(1)}_{fi}$ becomes
\begin{equation}
  \begin{split}
    S^{(1,I_1)}_{fi} = & \ (2\pi)^{-11/2}i\kappa^3\int d^3\vec{p}_1d^3\vec{q_1}d^3\vec{k}d^3\vec{p}_2d^3\vec{q}_2\hat{\psi}_1(\vec{p}_1)\hat{\psi}_2(\vec{q}_1)
                         \hat{\Psi}(\vec{p}_2,\vec{q}_2)\hat{\phi}^{\mu\nu}(\vec{k})\\
                       & \ \int d^4l_1\ldots d^4l_3\hat{g}(l_1)\ldots\hat{g}(l_3)b^{\al\be\ga\vep}\hat{D}^F_0(l_1+l_3+p_1-k-q_2)\\
                       & \ [-l_1^{\al}-p_1^{\al}+k^{\al}+q_2^{\al}][-l_1^{\si}-p_1^{\si}+k^{\si}+q_2^{\si}]\hat{D}^F_m(-l_1+k+q_2)\\
                       & \ b^{\mu\nu\ro\si}p_1^{\be}q_2^{\ro}p_2^{\ga}q_1^{\vep}\bigl(32\omega(\vec{k})\omega(\vec{p}_1)\omega(\vec{p}_2)\omega(\vec{q}_1)
                         \omega(\vec{q_2})\bigr)^{-1/2}\\
                       & \ \delta^{(4)}(l_1+l_2+l_3+p_1+q_1-q_2-p_2-k)\\
                     = & \ (2\pi)^{-15/2}i\kappa^3\int d^3\vec{p}_1d^3\vec{q_1}d^3\vec{k}d^3\vec{p}_2d^3\vec{q}_2\hat{\psi}_1(\vec{p}_1)\hat{\psi}_2(\vec{q}_1)
                         \hat{\Psi}(\vec{p}_2,\vec{q}_2)\hat{\phi}^{\mu\nu}(\vec{k})\\
                       & \ \int d^4l_1\ldots d^4l_3\hat{g}(l_1)\ldots\hat{g}(l_3)b^{\al\be\ga\vep}\frac{-l_1^{\al}-p_1^{\al}+k^{\al}+q_2^{\al}}{(l_1+l_3+p_1-k-q_2)^2+i0}
                         b^{\mu\nu\ro\si}\\
                       & \ \frac{-l_1^{\si}-p_1^{\si}+k^{\si}+q_2^{\si}}{m^2-(-l_1+k+q_2)^2-i0}p_1^{\be}q_2^{\ro}p_2^{\ga}q_1^{\vep}\bigl(32\omega(\vec{k})
                         \omega(\vec{p}_1)\omega(\vec{p}_2)\omega(\vec{q}_1)\omega(\vec{q_2})\bigr)^{-1/2}\\
                       & \ \delta^{(4)}(l_1+l_2+l_3+p_1+q_1-q_2-p_2-k)\\ 
  \end{split}
\label{Sfi4}
\end{equation}
where we have inserted the propagators explicitely. In this expression we can carry out the adiabatic limit in the variable $l_2$ since it doesn't appear in the argument of the propagators. We do this in the following way. First of all we choose a fixed testfunction $g_0\in \mathcal{S}(\mathbb{R}^4)$ with the property $g_0(0)=1$. Then a new testfunction is defined by
\begin{equation}
  g_{\vep}(x):=g_0(\vep x)
\end{equation}
The adiabatic limit $g_{\vep}\rightarrow 1$ is then equivalent to the limit $\vep\rightarrow 0$. The Fourier transform of $g_{\vep}$ is given by
\begin{equation}
  \hat{g}_{\vep}(p)=\frac{1}{\vep^4}\hat{g}_0\Bigl(\frac{p}{\vep}\Bigr)
\end{equation}
Then we have
\begin{equation}
  \lim_{\vep\rightarrow 0} \frac{1}{\vep^4}\hat{g}_0\Bigl(\frac{p}{\vep}\Bigr)=(2\pi)^2\delta^{(4)}(p)
\label{adia}
\end{equation}
Using this in (\ref{Sfi4}) the integration w.r.t. $l_2$ gives just the factor $(2\pi)^2$.

We are interested in the $S$-matrix element in the limit where only gravitons at low momenta (soft gravitons) are emited. So we omit all small quantities in the numerators, and  we take only the leading singularities in the denominators. This is known as the eikonal approximation in the literature, see e.g. \cite{css:qcd}. With this approximation we obtain for the argument of the massless propagator
\begin{equation}
  \begin{split}
    (p_1-q_2-k+l_1+l_3)^2 = & \ (p_1-q_2)^2+(-k+l_1+l_3)^2-2(p_1-q_2)(k-l_1-l_3)\\
                          = & \ (p_1-q_2)^2+o(\vep)\\
  \end{split}
\end{equation}
where $\vep$ is the scaling parameter from the adiabatic limit. The argument of the massive propagator becomes
\begin{equation}
  \begin{split}
    m^2-(-l_1+k+q_2)^2 = & \ m^2-q_2^2-(-l_1+k)^2+2q_2(l_1-k)\\
                       = & \ 2q_2(l_1-k)+o(\vep^2)\\
  \end{split}
\end{equation}
since the momentum $q_2$ is on the mass shell. Then the $S$-matrix element becomes
\begin{equation}
  \begin{split}
     S^{(1,I_1)}_{fi} = & \ (2\pi)^{-11/2}i\kappa^3\int d^3\vec{p}_1d^3\vec{q_1}d^3\vec{k}d^3\vec{p}_2d^3\vec{q}_2\hat{\psi}_1(\vec{p}_1)\hat{\psi}_2(\vec{q}_1)
                          \hat{\Psi}(\vec{p}_2,\vec{q}_2)\hat{\phi}^{\mu\nu}(\vec{k})\\
                        & \ \int d^4l_1d^4l_3\hat{g}(l_1)\hat{g}(l_3)b^{\al\be\ga\vep}\frac{(q_2^{\al}-p_1^{\al})}{(p_1-q_2)^2}\frac{(q_2^{\si}-p_1^{\si})}{2q_2(l_1-k)}
                          b^{\mu\nu\ro\si}\\
                        & \ p_1^{\be}q_2^{\ro}p_2^{\ga}q_1^{\vep}\bigl(32\omega(\vec{k})\omega(\vec{p}_1)\omega(\vec{p}_2)\omega(\vec{q}_1)\omega(\vec{q}_2)\bigr)^{-1/2}\\
                        & \ \delta^{(4)}(l_1+l_3+p_1+q_1-q_2-p_2-k)\\
  \end{split}
\label{Sfi5}
\end{equation}
Now the adiabatic limit in $l_3$ can be done as above and we arrive at the result for the $S$-matrix element
\begin{equation}
  \begin{split}
    S^{(1,I_1)}_{fi} = & \ (2\pi)^{-7/2}i\kappa^3\int d^3\vec{p}_1d^3\vec{q_1}d^3\vec{k}d^3\vec{p}_2d^3\vec{q}_2\hat{\psi}_1(\vec{p}_1)\hat{\psi}_2(\vec{q}_1)
                          \hat{\Psi}(\vec{p}_2,\vec{q}_2)\hat{\phi}^{\mu\nu}(\vec{k})\\
                       & \ \int d^4l_1\hat{g}(l_1)b^{\al\be\ga\vep}\frac{(q_2^{\al}-p_1^{\al})}{(p_1-q_2)^2}\frac{(q_2^{\si}-p_1^{\si})}{2q_2(l_1-k)}b^{\mu\nu\ro\si} \\
                       & \ p_1^{\be}q_2^{\ro}p_2^{\ga}q_1^{\vep}\bigl(32\omega(\vec{k})\omega(\vec{p}_1)\omega(\vec{p}_2)\omega(\vec{q}_1)\omega(\vec{q}_2)\bigr)^{-1/2}\\
                       & \ \delta^{(4)}(l_1+p_1+q_1-q_2-p_2-k)\\
  \end{split}
\label{Sfi6}
\end{equation}
There remains to calculate the other three parts $S^{(1,I_2)}_{fi}\ldots S^{(1,I_4)}_{fi}$. The calculations are similiar to the one above so we only present the results
\begin{equation}
  \begin{split}
    S^{(1,I_2)}_{fi} = & \ (2\pi)^{-7/2}i\kappa^3\int d^3\vec{p}_1d^3\vec{q_1}d^3\vec{k}d^3\vec{p}_2d^3\vec{q}_2\hat{\psi}_1(\vec{p}_1)\hat{\psi}_2(\vec{q}_1)
                          \hat{\Psi}(\vec{p}_2,\vec{q}_2)\hat{\phi}^{\mu\nu}(\vec{k})\\
                       & \ \int d^4l_1\hat{g}(l_1)b^{\al\be\ga\vep}\frac{(q_2^{\al}-q_1^{\al})}{(q_1-q_2)^2}\frac{(q_2^{\si}-q_1^{\si})}{2q_2(l_1-k)}b^{\mu\nu\ro\si} \\
                       & \ q_1^{\be}q_2^{\ro}p_2^{\ga}p_1^{\vep}\bigl(32\omega(\vec{k})\omega(\vec{p}_1)\omega(\vec{p}_2)\omega(\vec{q}_1)\omega(\vec{q}_2)\bigr)^{-1/2}\\
                       & \ \delta^{(4)}(l_1+q_1+p_1-q_2-p_2-k)\\
    S^{(1,I_3)}_{fi} = & \ (2\pi)^{-7/2}i\kappa^3\int d^3\vec{p}_1d^3\vec{q_1}d^3\vec{k}d^3\vec{p}_2d^3\vec{q}_2\hat{\psi}_1(\vec{p}_1)\hat{\psi}_2(\vec{q}_1)
                          \hat{\Psi}(\vec{p}_2,\vec{q}_2)\hat{\phi}^{\mu\nu}(\vec{k})\\
                       & \ \int d^4l_1\hat{g}(l_1)b^{\al\be\ga\vep}\frac{(p_2^{\al}-p_1^{\al})}{(p_1-p_2)^2}\frac{(p_2^{\si}-p_1^{\si})}{2p_2(l_1-k)}b^{\mu\nu\ro\si} \\
                       & \ p_1^{\be}p_2^{\ro}q_2^{\ga}q_1^{\vep}\bigl(32\omega(\vec{k})\omega(\vec{p}_1)\omega(\vec{p}_2)\omega(\vec{q}_1)\omega(\vec{q}_2)\bigr)^{-1/2}\\
                       & \ \delta^{(4)}(l_1+q_1+p_1-q_2-p_2-k)\\
    S^{(1,I_4)}_{fi} = & \ (2\pi)^{-7/2}i\kappa^3\int d^3\vec{p}_1d^3\vec{q_1}d^3\vec{k}d^3\vec{p}_2d^3\vec{q}_2\hat{\psi}_1(\vec{p}_1)\hat{\psi}_2(\vec{q}_1)
                          \hat{\Psi}(\vec{p}_2,\vec{q}_2)\hat{\phi}^{\mu\nu}(\vec{k})\\
                       & \ \int d^4l_1\hat{g}(l_1)b^{\al\be\ga\vep}\frac{(p_2^{\al}-q_1^{\al})}{(q_1-p_2)^2}\frac{(p_2^{\si}-q_1^{\si})}{2p_2(l_1-k)}b^{\mu\nu\ro\si} \\
                       & \ q_1^{\be}p_2^{\ro}q_2^{\ga}p_1^{\vep}\bigl(32\omega(\vec{k})\omega(\vec{p}_1)\omega(\vec{p}_2)\omega(\vec{q}_1)\omega(\vec{q}_2)\bigr)^{-1/2}\\
                       & \ \delta^{(4)}(l_1+q_1+p_1-q_2-p_2-k)\\ 
  \end{split}
\end{equation}
Now we are ready to consider the differential cross section. This will be done in detail in the next subsection. 

%#*#*#*#*#*#*#*#*#*#*#*#*#*#*#*#*#*#*#*#*#*#*#*#*#*#*#*#*#*#*#*#*#*#*#*#*#*#*#*#*#*#*#*#*#*#*#*#*#*#*#*#*#*#*#*#*#*#*#*#*#*#*#*#*#*#*#*#*#*#*#*#*#*#*#
\subsection{Towards the differential cross section--Step 2: Calculation of  $|S_{fi}|^2$}
We have to calculate the absolute square of the $S$-matrix element $S^{(1)}_{fi}$. We do this again separately for the four parts $S^{(1,I_1)}_{fi}\ldots S^{(1,I_4)}_{fi}$. For the first part we start with (\ref{Sfi6}) and write
\begin{equation}
  \begin{split}
    \big\lvert S^{(1,I_1)}_{fi}\big\rvert^2 = & \ (2\pi)^{-7}\kappa^6\int d^3\vec{p}_1d^3\vec{q}_1d^3\vec{k}d^3\vec{p}_2d^3\vec{q}_2d^3\vec{p_1}^{\prime}d^3\vec{q_1}^{\prime}
                                                d^3\vec{k}^{\prime}d^3\vec{p_2}^{\prime}d^3\vec{q_2}^{\prime}\\
                                              & \ \hat{\psi}_1(\vec{p}_1)\hat{\psi}_2(\vec{q_1})\hat{\Psi}(\vec{p}_2,\vec{q}_2)\hat{\phi}^{\mu\nu}(\vec{k})
                                                \hat{\psi}_1(\vec{p_1}^{\prime})^{\ast}\hat{\psi}_2(\vec{q_1}^{\prime})^{\ast}\hat{\Psi}(\vec{p_2}^{\prime},
                                                \vec{q_2}^{\prime})^{\ast}\hat{\phi}^{\mu^{\prime}\nu^{\prime}}(\vec{k}^{\prime})^{\ast}\\
                                              & \ \int d^4l_1d^4l_2\hat{g}(l_1)\hat{g}(l_2)^{\ast}b^{\al\be\ga\vep}\frac{(q_2^{\al}-p_1^{\al})}{(q_2-p_1)^2}
                                                \frac{(q_2^{\si}-p_1^{\si})}{2q_2(l_1-k)}b^{\mu\nu\ro\si}\\
                                              & \ p_1^{\be}q_2^{\ro}p_2^{\ga}q_1^{\vep}b^{\al^{\prime}\be^{\prime}\ga^{\prime}\vep^{\prime}}\frac{({q_2^{\prime}}^{\al^{\prime}}-
                                                {p_1^{\prime}}^{\al^{\prime}})}{(q_2^{\prime}-p_1^{\prime})^2}\frac{({q_2^{\prime}}^{\si^{\prime}}
                                                -{p_1^{\prime}}^{\si^{\prime}})}{2q_2^{\prime}(l_2-k^{\prime})}b^{\mu^{\prime}\nu^{\prime}\ro^{\prime}\si^{\prime}}\\
                                              & \ {p_1^{\prime}}^{\be^{\prime}}{q_2^{\prime}}^{\ro^{\prime}}{p_2^{\prime}}^{\ga^{\prime}}{q_1^{\prime}}^{\vep^{\prime}}
                                                \bigl(32\omega(\vec{k})\omega(\vec{p}_1)\omega(\vec{p}_2)\omega(\vec{q}_1)\omega(\vec{q}_2)\bigr)^{-1/2}\\
                                              & \ \bigl(32\omega(\vec{k^{\prime}})\omega(\vec{p_1}^{\prime})\omega(\vec{p_2}^{\prime})\omega(\vec{q_1}^{\prime})
                                                \omega(\vec{q_2}^{\prime})\bigr)^{-1/2}\\
                                              & \ \delta^{(4)}(l_1+p_1+q_1-q_2-p_2-k)\delta^{(4)}(l_2+p_1^{\prime}+q_1^{\prime}-q_2^{\prime}-p_2^{\prime}-k^{\prime})\\
  \end{split}
\label{sigma1}
\end{equation}
where a $^{\ast}$ indicates complex conjugation. We use the completeness relations of the final states, i.e.
\begin{align}
  \sum_f \hat{\Psi}(\vec{p_2},\vec{q_2})\hat{\Psi}(\vec{p_2}^{\prime},\vec{q_2}^{\prime})^{\ast} & =\delta^{(3)}(\vec{p_2}-\vec{p_2}^{\prime})
                                                                                                  \delta^{(3)}(\vec{q_2}-\vec{q_2}^{\prime})\\
  \sum_f \hat{\phi}^{\mu\nu}(\vec{k})\hat{\phi}^{\mu^{\prime}\nu^{\prime}}(\vec{k}^{\prime})^{\ast} & =\eta^{\mu\nu}\eta^{\mu^{\prime}\nu^{\prime}}
                                                                                                     \delta^{(3)}(\vec{k}-\vec{k}^{\prime})
\end{align}
where the sum is taken over all final states. Then we have
\begin{equation}
  \begin{split}
    \sum_f \big\lvert S^{(1,I_1)}_{fi}\big\rvert^2 = & \ (2\pi)^{-7}\kappa^6\int d^3\vec{p}_1d^3\vec{q}_1d^3\vec{k}d^3\vec{p}_2d^3\vec{q}_2d^3\vec{p_1}^{\prime}d^3
                                                    \vec{q_1}^{\prime}\hat{\psi}_1(\vec{p}_1)\hat{\psi}_2(\vec{q_1})\\
                                                  & \ \hat{\psi}_1(\vec{p_1}^{\prime})^{\ast}\hat{\psi}_2(\vec{q_1}^{\prime})^{\ast}\int d^4l_1d^4l_2\hat{g}(l_1)
                                                    \hat{g}(l_2)^{\ast}b^{\al\be\ga\vep}\frac{(q_2^{\al}-p_1^{\al})}{(q_2-p_1)^2}\frac{(q_2^{\ro}-p_1^{\ro})}{2q_2(l_1-k)}\\
                                                  & \ p_1^{\be}q_2^{\ro}p_2^{\ga}q_1^{\vep}b^{\al^{\prime}\be^{\prime}\ga^{\prime}\vep^{\prime}}\frac{(q_2^{\al^{\prime}}-
                                                    {p_1^{\prime}}^{\al^{\prime}})}{(q_2-p_1^{\prime})^2}\frac{(q_2^{\ro^{\prime}}-{p_1^{\prime}}^{\ro^{\prime}})}
                                                    {2q_2(l_2-k)}{p_1^{\prime}}^{\be^{\prime}}q_2^{\ro^{\prime}}p_2^{\ga^{\prime}}{q_1^{\prime}}^{\vep^{\prime}}\\
                                                  & \ \frac{\delta^{(4)}(l_1+p_1+q_1-q_2-p_2-k)\delta^{(4)}(l_2+p_1^{\prime}+q_1^{\prime}-q_2-p_2-k)}
                                                    {\sqrt{\bigl(32\omega(\vec{k})\omega(\vec{p}_1)\omega(\vec{p}_2)\omega(\vec{q}_1)\omega(\vec{q}_2)\bigr)\bigl(32
                                                    \omega(\vec{k})\omega(\vec{p_1}^{\prime})\omega(\vec{p_2})\omega(\vec{q_1}^{\prime})\omega(\vec{q_2})\bigr)}}\\
  \end{split}
\label{sigme2}
\end{equation}
Now we assume that the wavefunctions $\hat{\psi}_1$ and $\hat{\psi}_2$ are sharply concentrated around the initial momenta $p_i$ and $q_i$, so that we can replace the momenta in the propagators by these initial momenta. Then the expression can be simplified to
\begin{equation}
  \begin{split}
    \sum_f \big\lvert S^{(1,I_1)}_{fi}\big\rvert^2 = & \ (2\pi)^{-7}\kappa^6\int d^4l_1d^4l_2\hat{g}(l_1)\hat{g}(l_2)^{\ast}\int d^3\vec{k}d^3\vec{p_2}d^3\vec{q_2}
                                                    b^{\al\be\ga\vep}\\
                                                  & \ \frac{(q_2^{\al}-p_i^{\al})}{(q_2-p_i)^2}\frac{(q_2^{\ro}-p_i^{\ro})}{2q_2(l_1-k)}p_i^{\be}q_2^{\ro}p_2^{\ga}q_i^{\vep}
                                                    b^{\al^{\prime}\be^{\prime}\ga^{\prime}\vep^{\prime}}\frac{(q_2^{\al^{\prime}}-p_i^{\al^{\prime}})}{(q_2-p_i)^2}
                                                    \frac{(q_2^{\ro^{\prime}}-p_i^{\ro^{\prime}})}{2q_2(l_2-k)}\\
                                                  & \ p_i^{\be^{\prime}}q_2^{\ro^{\prime}}p_2^{\ga^{\prime}}q_i^{\vep^{\prime}}\bigl[32\omega(\vec{k})\omega(\vec{p_i})
                                                    \omega(\vec{p_2})\omega(\vec{q_i})\omega(\vec{q_2})\bigr]^{-1}\\
                                                  & \ \int d^3\vec{p}_1d^3\vec{q}_1d^3\vec{p_1}^{\prime}d^3\vec{q_1}^{\prime}\hat{\psi}_1(\vec{p}_1)\hat{\psi}_2(\vec{q_1})
                                                    \hat{\psi}_1(\vec{p_1}^{\prime})^{\ast}\hat{\psi}_2(\vec{q_1}^{\prime})^{\ast}\\
                                                  & \ \delta^{(4)}(l_1+p_1+q_1-q_2-p_2-k)\delta^{(4)}(l_2+p_1^{\prime}+q_1^{\prime}-q_2-p_2-k)\\
  \end{split}
\label{sigma3}
\end{equation}
We observe that the last integral herein depends on the initial state only. We denote it by $F(P,l_1,l_2)$, where we have introduced $P=p_2+q_2+k$. We have
\begin{equation}
  \begin{split}
    F(P,l_1,l_2) = & \ \int d^3\vec{p}_1d^3\vec{q}_1d^3\vec{p_1}^{\prime}d^3\vec{q_1}^{\prime}\hat{\psi}_1(\vec{p}_1)\hat{\psi}_2(\vec{q_1})
                     \hat{\psi}_1(\vec{p_1}^{\prime})^{\ast}\hat{\psi}_2(\vec{q_1}^{\prime})^{\ast}\\
                   & \ \delta^{(4)}(l_1+p_1+q_1-q_2-p_2-k)\delta^{(4)}(l_2+p_1^{\prime}+q_1^{\prime}-q_2-p_2-k)\\
                 = & \ (2\pi)^{-8}\int d^4y_1d^4y_2d^3\vec{p}_1\ldots d^3\vec{q_1}^{\prime}\hat{\psi}_1(\vec{p}_1)\hat{\psi}_2(\vec{q_1})
                     \hat{\psi}_1(\vec{p_1}^{\prime})^{\ast}\hat{\psi}_2(\vec{q_1}^{\prime})^{\ast}\\
                   & \ \exp\bigl[-i(l_1+p_1+q_1-P)y_1\bigr]\exp\bigl[+i(l_2+p_1^{\prime}+q_1^{\prime}-P)y_2\bigr]\\
                 = & \ \int d^4y_1d^4y_2\psi_1(y_1)\psi_2(y_1)\psi_1(y_2)^{\ast}\psi_2(y_2)^{\ast}\exp[-il_1y_1]\exp[+il_2y_2]\\
                   & \ \exp\bigl[-i(y_1-y_2)P\bigr]\\
  \end{split}
\end{equation}
The function $F$ is normalized according to
\begin{equation}
  \int F(P,l_1,l_2)d^4P=(2\pi)^2\int d^4y|\psi_1(y)|^2|\psi_2(y)|^2\exp\bigl[-i(l_1-l_2)y\bigr]
\end{equation}
and it is concentrated around $P=p_2+q_2+k\approx p_i+q_i$. In the limit of infintely sharp wave packets we may represent it by
\begin{equation}
  F(P,l_1,l_2)=(2\pi)^2\delta^{(4)}(P-p_i-q_i)\int d^4y|\psi_1(y)|^2|\psi_2(y)|^2\exp\bigl[-i(l_1-l_2)y\bigr]
\end{equation}
This will be inserted in (\ref{sigma3}) and we obtain
\begin{equation} 
  \begin{split}
    \sum_f \big\lvert S^{(1,I_1)}_{fi}\big\rvert^2 = & \ (2\pi)^{-5}\kappa^6\int d^4l_1d^4l_2\hat{g}(l_1)\hat{g}(l_2)^{\ast}\int d^3\vec{k}d^3\vec{p_2}d^3\vec{q_2}
                                                    b^{\al\be\ga\vep}\\
                                                  & \ \frac{(q_2^{\al}-p_i^{\al})}{(q_2-p_i)^2}\frac{(q_2^{\ro}-p_i^{\ro})}{2q_2(l_1-k)}p_i^{\be}q_2^{\ro}p_2^{\ga}q_i^{\vep}
                                                    b^{\al^{\prime}\be^{\prime}\ga^{\prime}\vep^{\prime}}\frac{(q_2^{\al^{\prime}}-p_i^{\al^{\prime}})}{(q_2-p_i)^2}
                                                    \frac{(q_2^{\ro^{\prime}}-p_i^{\ro^{\prime}})}{2q_2(l_2-k)}\\
                                                  & \ p_i^{\be^{\prime}}q_2^{\ro^{\prime}}p_2^{\ga^{\prime}}q_i^{\vep^{\prime}}\bigl[32\omega(\vec{k})\omega(\vec{p_i})
                                                    \omega(\vec{p_2})\omega(\vec{q_i})\omega(\vec{q_2})\bigr]^{-1}\\
                                                  & \ \delta^{(4)}(p_2+q_2+k-p_i-q_i)\int d^4y|\psi_1(y)|^2|\psi_2(y)|^2\exp\bigl[-i(l_1-l_2)y\bigr]\\
  \end{split}
\label{sigma4}
\end{equation}
For the definition of the cross section we follow sect. 3.4 of \cite{sch:qed}. We can set $l_1=l_2$ in the last integral because we neglect all contributions $o(\vep)$ in the numerator, see the remarks after (\ref{Sfi4}). With the notation as in \cite{sch:qed} we define the cross section as
\begin{equation}
  \sigma=\lim_{R\rightarrow\infty}\pi R^2\sum_f|S_{fi}(R)|^2
\end{equation}
where $R$ is the radius of the beam of incoming particles. Then we restrict ourself again to (\ref{T3Term1}) of $T_3$ and furthermore to the first integral $I_1$ (\ref{I1}) of $S_{fi}$. If we denote the integration variables $p_2$ and $q_2$ by $p_f$ and $q_f$, respectively, then we obtain for the cross section
\begin{equation}
  \begin{split}
    \sigma = & \ (2\pi)^{-5}\kappa^6\bigl[4\omega(\vec{p_i})\omega(\vec{q_i})\bigr]^{-1}\frac{\omega(\vec{p_i})\omega(\vec{q_i})}{\sqrt{\bigl((p_iq_i)^2-m^4\bigr)}}\int 
               d^4l_1d^4l_2\hat{g}(l_1)\hat{g}(l_2)^{\ast}\\
             & \ \int\frac{d^3\vec{k}}{2\omega(\vec{k})}\frac{d^3\vec{q_f}}{2\omega(\vec{q_f})}\frac{d^3\vec{p_f}}{2\omega(\vec{p_f})}b^{\al\be\ga\vep}\frac{(q_f^{\al}-
               p_i^{\al})}{(q_f-p_i)^2}\frac{(q_f^{\ro}-p_i^{\ro})}{2q_f(l_1-k)}p_i^{\be}q_f^{\ro}p_f^{\ga}q_i^{\vep}b^{\al^{\prime}\be^{\prime}\ga^{\prime}\vep^{\prime}}\\
             & \ \frac{(q_f^{\al^{\prime}}-p_i^{\al^{\prime}})}{(q_f-p_i)^2}\frac{(q_f^{\ro^{\prime}}-p_i^{\ro^{\prime}})}{2q_f(l_2-k)}p_i^{\be^{\prime}}q_f^{\ro^{\prime}}
               p_f^{\ga^{\prime}}q_i^{\vep^{\prime}}\delta^{(3)}(\vec{p_f}+\vec{q_f}+\vec{k}-\vec{p_i}-\vec{q_i})\\
             & \ \delta(\omega(\vec{p_f})+\omega(\vec{q_f})+\omega(\vec{k})-\omega(\vec{p_i})-\omega(\vec{q_i}))\\
  \end{split}
\end{equation}
To carry out the remainig integrations we have to know the integrand explicitely. So at this point of the calculation the tensor structure of the integrand becomes important and we have to write down the various contractions between the four-vectors $p_i,q_i,p_f,q_f$. With the tensor $b^{\mu\nu\ro\si}$ (\ref{b-tensor}) we get for the cross section 
\begin{equation}
  \begin{split}
    \sigma = & \ \frac{1}{4}\frac{(2\pi)^{-5}\kappa^6}{\sqrt{\bigl((p_iq_i)^2-m^4\bigr)}}\int d^4l_1d^4l_2\hat{g}(l_1)\hat{g}(l_2)^{\ast}\int\frac{d^3\vec{k}}
               {2\om(\vec{k})}\frac{d^3\vec{q_f}}{2\om(\vec{q_f})}\frac{d^3\vec{p_f}}{2\om(\vec{p_f})}\\
             & \ \Biggl[\Bigl[\bigl(\om(\vec{q_f})\om(\vec{p_f})-\vec{q_f}\vec{p_f}\bigr)(p_iq_i)+\bigl(\om(\vec{q_f})\om(\vec{q_i})-\vec{q_f}\vec{q_i}\bigr)\bigl(\om(\vec{p_i})
               \om(\vec{p_f})-\vec{p_i}\vec{p_f}\bigr)\\
             & -\bigl(\om(\vec{q_f})\om(\vec{p_i})-\vec{q_f}\vec{p_i}\bigr)\bigl(\om(\vec{p_f})\om(\vec{q_i})-\vec{p_f}\vec{q_i}\bigr)-\bigl(\om(\vec{p_i})\om(\vec{p_f})-
               \vec{p_i}\vec{p_f}\bigr)(p_iq_i)\\
             & -(p_iq_i)\bigl(\om(\vec{p_i})\om(\vec{p_f})-\vec{p_i}\vec{p_f}\bigr)+p_i^2\bigl(\om(\vec{p_f})\om(\vec{q_i})-\vec{p_f}\vec{q_i}\bigr)\Bigr]\bigl(\om(\vec{q_f})^2
               -\vec{q_f}^2\bigr)\\
             & -\Bigl[\bigl(\om(\vec{q_f})\om(\vec{p_f})-\vec{q_f}\vec{p_f}\bigr)(p_iq_i)-\bigl(\om(\vec{q_f})\om(\vec{q_i})-\vec{q_f}\vec{q_i}\bigr)\bigl(\om(\vec{p_i})
               \om(\vec{p_f})-\vec{p_i}\vec{p_f}\bigr)\\
             & +\bigl(\om(\vec{q_f})\om(\vec{p_i})-\vec{q_f}\vec{p_i}\bigr)\bigl(\om(\vec{p_f})\om(\vec{q_i})-\vec{p_f}\vec{q_i}\bigr)+\bigl(\om(\vec{p_i})\om(\vec{p_f})-
               \vec{p_i}\vec{p_f}\bigr)(p_iq_i)\\
             & +(p_iq_i)\bigl(\om(\vec{p_i})\om(\vec{p_f})-\vec{p_i}\vec{p_f}\bigr)-p_i^2\bigl(\om(\vec{p_f})\om(\vec{q_i})-\vec{p_f}\vec{q_i}\bigr)\Bigr]\bigl(\om(\vec{q_f})
               \om(\vec{p_i})-\vec{q_f}\vec{p_i}\bigr)\Biggr]^2\\
             & \ \frac{\delta^{(3)}(\vec{p_f}+\vec{q_f}+\vec{k}-\vec{p_i}-\vec{q_i})\delta(\omega(\vec{p_f})+\omega(\vec{q_f})+\omega(\vec{k})-\omega(\vec{p_i})
               -\omega(\vec{q_i}))}{(q_f-p_i)^22q_f(l_1-k)(q_f-p_i)^22q_f(l_2-k)}\\
  \end{split}
\end{equation}
It is convenient to choose the center-of-mass system, defined by $\vec{p_i}+\vec{q_i}=0$. We can easily do the integration w.r.t. $\vec{p_f}$:
\begin{equation}
  \begin{split}
    \sigma = & \ \frac{1}{4}\frac{(2\pi)^{-5}\kappa^6}{\sqrt{\bigl((p_iq_i)^2-m^4\bigr)}}\int d^4l_1d^4l_2\hat{g}(l_1)\hat{g}(l_2)^{\ast}\int\frac{d^3\vec{k}}
               {2\om(\vec{k})}\frac{d^3\vec{q_f}}{2\om(\vec{q_f})}\frac{1}{2\om(\vec{q_f}+\vec{k})}\\
             & \ \Biggl[\Bigl[\bigl(\om(\vec{q_f})\om(\vec{q_f}+\vec{k})+\vec{q_f}(\vec{q_f}+\vec{k})\bigr)(p_iq_i)+\bigl(\om(\vec{q_f})\om(\vec{q_i})-\vec{q_f}\vec{q_i}\bigr)
               \bigl(\om(\vec{p_i})\om(\vec{q_f}+\vec{k})\\
             & +\vec{p_i}(\vec{q_f}+\vec{k})\bigr)-\bigl(\om(\vec{q_f})\om(\vec{p_i})-\vec{q_f}\vec{p_i}\bigr)\bigl(\om(\vec{q_f}+\vec{k})\om(\vec{q_i})+(\vec{q_f}
               +\vec{k})\vec{q_i}\bigr)\\
             & -\bigl(\om(\vec{p_i})\om(\vec{q_f}+\vec{k})+\vec{p_i}(\vec{q_f}+\vec{k})\bigr)(p_iq_i)-(p_iq_i)\bigl(\om(\vec{p_i})\om(\vec{q_f}+\vec{k})+\vec{p_i}
               (\vec{q_f}+\vec{k})\bigr)\\
             & +\vec{p_i}^2\bigl(\om(\vec{q_f}+\vec{k})\om(\vec{q_i})+(\vec{q_f}+\vec{k})\vec{q_i}\bigr)\Bigr]\bigl(\om(\vec{q_f})^2-\vec{q_f}^2\bigr)
               -\Bigl[\bigl(\om(\vec{q_f})\om(\vec{q_f}+\vec{k})\\
             & +\vec{q_f}(\vec{q_f}+\vec{k})\bigr)(p_iq_i)-\bigl(\om(\vec{q_f})\om(\vec{q_i})-\vec{q_f}\vec{q_i}\bigr)\bigl(\om(\vec{p_i})\om(\vec{q_f}+\vec{k})
               +\vec{p_i}(\vec{q_f}+\vec{k})\bigr)\\
             & +\bigl(\om(\vec{q_f})\om(\vec{p_i})-\vec{q_f}\vec{p_i}\bigr)\bigl(\om(\vec{q_f}+\vec{k})\om(\vec{q_i})+(\vec{q_f}+\vec{k})\vec{q_i}\bigr)+\bigl(\om(\vec{p_i})
               \om(\vec{q_f}+\vec{k})\\
             & +\vec{p_i}(\vec{q_f}+\vec{k})\bigr)(p_iq_i)+(p_iq_i)\bigl(\om(\vec{p_i})\om(\vec{q_f}+\vec{k})+\vec{p_i}(\vec{q_f}+\vec{k})\bigr)\\
             & -p_i^2\bigl(\om(\vec{q_f}+\vec{k})\om(\vec{q_i})+(\vec{q_f}+\vec{k})\vec{q_i}\bigr)\Bigr]\bigl(\om(\vec{q_f})\om(\vec{p_i})-\vec{q_f}\vec{p_i}\bigr)\Biggr]^2\\
             & \ \frac{\delta(\om(\vec{q_f})+\om(\vec{q_f}+\vec{k})+\om(\vec{k})-2\om(\vec{p_i}))}{(q_f-p_i)^22q_f(l_1-k)(q_f-p_i)^22q_f(l_2-k)}\\
  \end{split}
\end{equation}
Now we are left with the integration w.r.t. $\vec{q_f}$. In order to be able to do this we have to rewrite the argument of the delta distribution showing the dependence on 
$|\vec{q_f}|$ explicitely. We want to use the formula
\begin{equation}
  \delta(f(x))=\sum_{i=1}^n\frac{1}{|f^{\prime}(x_i)|}\delta(x-x_i)
\end{equation}
where the $x_i$ are simple zeros of the function $f$. For that purpose we define
\begin{equation}
  \begin{split}
    f(\abs{q_f}) = & \ \sqrt{\vec{q_f}^2+m^2}+\sqrt{(\vec{q_f}+\vec{k})^2+m^2}+\om(\vec{k})-2\om(\vec{p_i})\\
                   = & \ \sqrt{\abs{q_f}^2+m^2}+\sqrt{\abs{q_f}^2+\abs{k}^2+2\abs{q_f}\abs{k}\cos\theta+m^2}+\abs{k}-2\om(\vec{p_i})\\
  \end{split}
\end{equation}
where $\theta=\sphericalangle (\vec{q_f},\vec{k})$ and we have used the notation $\abs{k}=|\vec{k}|,\,\abs{q_f}=|\vec{q_f}|$. In order to simplify the notation in the following formulas we write $\om=2\om(\vec{p_i})$. The zeros of the function $f$ are given by
\begin{equation}
  \begin{split}
    \abs{q_f}^{(1)}(\abs{k},\theta) = & \ A(\abs{k},\theta)+B(\abs{k},\theta)\\
    \abs{q_f}^{(2)}(\abs{k},\theta) = & \ A(\abs{k},\theta)-B(\abs{k},\theta)\\
  \end{split}
\label{zeros}
\end{equation}
where we have introduced
\begin{equation}
  \begin{split}
    A(\abs{k},\theta) = & \ \frac{\om(2\abs{k}-\om)\abs{k}\cos\theta}{2(\om-\abs{k})^2-2\abs{k}^2\cos^2\theta} \\
    B(\abs{k},\theta) = & \ \frac{(\om-\abs{k})\sqrt{\om^4-4\om^3\abs{k}+8\om\abs{k}m^2-2\abs{k}^2m^2+4\om^2(\abs{k}^2-m^2)+2\abs{k}^2m^2\cos 2\theta}}
                          {2(\om-\abs{k})^2-2\abs{k}^2\cos^2\theta} \\
  \end{split}
\end{equation}
Clearly these solutions are itself functions of the momentum $\abs{k}$ and the angle $\theta$. If we consider the limit of these functions (\ref{zeros}) as $\abs{k}$ goes to zero we obtain
\begin{equation}
  \lim_{\abs{k}\rightarrow 0}\abs{q_f}^{(j)}=\pm\abs{p_i},\ j=1,2    
\label{limitqf}
\end{equation}
where $\abs{p_i}=|\vec{p_i}|$. The derivative of $f$ with respect to $\abs{q_f}$ is given by
\begin{equation}
  \frac{df}{d\abs{q_f}}=\frac{\abs{q_f}}{\sqrt{m^2+\abs{q_f}^2}}+\frac{\abs{q_f}+\abs{k}\cos\theta}{\sqrt{\abs{k}^2+m^2+\abs{q_f}^2+2\abs{k}\abs{q_f}\cos\theta}}
\label{fprime}
\end{equation}
By inserting the two zeros of $f$ into (\ref{fprime}) we observe that, by taking the limit $\abs{k}\rightarrow 0$, we get a finite result, namely
\begin{equation}
  \lim_{\abs{k}\rightarrow 0}\frac{df(\abs{q_f})}{d\abs{q_f}}\bigg{\rvert}_{\abs{q_f}=\abs{q_f}^{(j)}}=\pm\frac{2\abs{p_i}}{\om(\vec{p_i})},\ j=1,2
\end{equation}
Now we can rewrite the cross section as
\begin{equation}
  \begin{split}
    \sigma = & \ \frac{1}{4}\frac{(2\pi)^{-5}\kappa^6}{\sqrt{\bigl((p_iq_i)^2-m^4\bigr)}}\int d^4l_1d^4l_2\hat{g}(l_1)\hat{g}(l_2)^{\ast}\int\frac{d^3\vec{k}}
               {2\om(\vec{k})}\frac{\abs{q_f}^2d\abs{q_f}d\Omega}{2\om(\vec{q_f})}\frac{1}{2\om(\vec{q_f}+\vec{k})}\\
             & \ \Biggl[\Bigl[\bigl(\om(\vec{q_f})\om(\vec{q_f}+\vec{k})+\vec{q_f}(\vec{q_f}+\vec{k})\bigr)(p_iq_i)+\bigl(\om(\vec{q_f})\om(\vec{q_i})-\vec{q_f}\vec{q_i}\bigr)
               \bigl(\om(\vec{p_i})\om(\vec{q_f}+\vec{k})\\
             & +\vec{p_i}(\vec{q_f}+\vec{k})\bigr)-\bigl(\om(\vec{q_f})\om(\vec{p_i})-\vec{q_f}\vec{p_i}\bigr)\bigl(\om(\vec{q_f}+\vec{k})\om(\vec{q_i})+(\vec{q_f}
               +\vec{k})\vec{q_i}\bigr)\\
             & -\bigl(\om(\vec{p_i})\om(\vec{q_f}+\vec{k})+\vec{p_i}(\vec{q_f}+\vec{k})\bigr)(p_iq_i)-(p_iq_i)\bigl(\om(\vec{p_i})\om(\vec{q_f}+\vec{k})+\vec{p_i}
               (\vec{q_f}+\vec{k})\bigr)\\
             & +\vec{p_i}^2\bigl(\om(\vec{q_f}+\vec{k})\om(\vec{q_i})+(\vec{q_f}+\vec{k})\vec{q_i}\bigr)\Bigr]\bigl(\om(\vec{q_f})^2-\vec{q_f}^2\bigr)
               -\Bigl[\bigl(\om(\vec{q_f})\om(\vec{q_f}+\vec{k})\\
             & +\vec{q_f}(\vec{q_f}+\vec{k})\bigr)(p_iq_i)-\bigl(\om(\vec{q_f})\om(\vec{q_i})-\vec{q_f}\vec{q_i}\bigr)\bigl(\om(\vec{p_i})\om(\vec{q_f}+\vec{k})
               +\vec{p_i}(\vec{q_f}+\vec{k})\bigr)\\
             & +\bigl(\om(\vec{q_f})\om(\vec{p_i})-\vec{q_f}\vec{p_i}\bigr)\bigl(\om(\vec{q_f}+\vec{k})\om(\vec{q_i})+(\vec{q_f}+\vec{k})\vec{q_i}\bigr)+\bigl(\om(\vec{p_i})
               \om(\vec{q_f}+\vec{k})\\
             & +\vec{p_i}(\vec{q_f}+\vec{k})\bigr)(p_iq_i)+(p_iq_i)\bigl(\om(\vec{p_i})\om(\vec{q_f}+\vec{k})+\vec{p_i}(\vec{q_f}+\vec{k})\bigr)\\
             & -p_i^2\bigl(\om(\vec{q_f}+\vec{k})\om(\vec{q_i})+(\vec{q_f}+\vec{k})\vec{q_i}\bigr)\Bigr]\bigl(\om(\vec{q_f})\om(\vec{p_i})-\vec{q_f}\vec{p_i}\bigr)\Biggr]^2\\ 
             & \ \frac{1}{(q_f-p_i)^22q_f(l_1-k)(q_f-p_i)^22q_f(l_2-k)}\sum_{j=1}^{2}\frac{1}{\lvert f^{\prime}(\abs{q_f})\rvert_{\abs{q_f}=\abs{q_f}^{(j)}}\rvert}
               \delta(\abs{q_f}-\abs{q_f}^{(j)})\\
  \end{split}
\end{equation}
Before we do the integration w.r.t. $\abs{q_f}$ let us denote by $F_1(\abs{q_f},\abs{k},\abs{k}\cos\theta)$ the expression which comes from the tensor part in the numerator, i.e.
\begin{equation}
  \begin{split}
    F_1(\abs{q_f},\abs{k},\abs{k}\cos\theta) = & \ \Biggl[\Bigl[\bigl(\om(\vec{q_f})\om(\vec{q_f}+\vec{k})+\vec{q_f}(\vec{q_f}+\vec{k})\bigr)(p_iq_i)+\bigl(\om(\vec{q_f})
                                                 \om(\vec{q_i})-\vec{q_f}\vec{q_i}\bigr)\\
                                               & \ \bigl(\om(\vec{p_i})\om(\vec{q_f}+\vec{k})+\vec{p_i}(\vec{q_f}+\vec{k})\bigr)-\bigl(\om(\vec{q_f})\om(\vec{p_i})
                                                 -\vec{q_f}\vec{p_i}\bigr)\bigl(\om(\vec{q_f}+\vec{k})\\
                                               & \ \om(\vec{q_i})+(\vec{q_f}+\vec{k})\vec{q_i}\bigr)-\bigl(\om(\vec{p_i})\om(\vec{q_f}+\vec{k})+\vec{p_i}(\vec{q_f}
                                                 +\vec{k})\bigr)(p_iq_i)-(p_iq_i)\\
                                               & \bigl(\om(\vec{p_i})\om(\vec{q_f}+\vec{k})+\vec{p_i}(\vec{q_f}+\vec{k})\bigr)+\vec{p_i}^2\bigl(\om(\vec{q_f}
                                                 +\vec{k})\om(\vec{q_i})+(\vec{q_f}+\vec{k})\vec{q_i}\bigr)\Bigr]\\
                                               & \ \bigl(\om(\vec{q_f})^2-\vec{q_f}^2\bigr)-\Bigl[\bigl(\om(\vec{q_f})\om(\vec{q_f}+\vec{k})+\vec{q_f}(\vec{q_f}+\vec{k})\bigr)
                                                 \ (p_iq_i)-\bigl(\om(\vec{q_f})\\
                                               & \ \om(\vec{q_i})-\vec{q_f}\vec{q_i}\bigr)\bigl(\om(\vec{p_i})\om(\vec{q_f}+\vec{k})+\vec{p_i}(\vec{q_f}+\vec{k})\bigr)
                                                 +\bigl(\om(\vec{q_f})\om(\vec{p_i})-\vec{q_f}\vec{p_i}\bigr)\\
                                               & \ \bigl(\om(\vec{q_f}+\vec{k})\om(\vec{q_i})+(\vec{q_f}+\vec{k})\vec{q_i}\bigr)+\bigl(\om(\vec{p_i})\om(\vec{q_f}+\vec{k})
                                                 +\vec{p_i}(\vec{q_f}+\vec{k})\bigr)\\
                                               & \ (p_iq_i)+(p_iq_i)\bigl(\om(\vec{p_i})\om(\vec{q_f}+\vec{k})+\vec{p_i}(\vec{q_f}+\vec{k})\bigr)-p_i^2\bigl(\om(\vec{q_f}
                                                 +\vec{k})\\
                                               & \ \om(\vec{q_i})+(\vec{q_f}+\vec{k})\vec{q_i}\bigr)\Bigr]\bigl(\om(\vec{q_f})\om(\vec{p_i})-\vec{q_f}\vec{p_i}\bigr)\Biggr]^2\\
  \end{split}
\end{equation}
Then the expression for the cross section is given by 
\begin{equation}
  \begin{split}
    \sigma = & \ \frac{1}{4}\frac{(2\pi)^{-5}\kappa^6}{\sqrt{\bigl((p_iq_i)^2-m^4\bigr)}}\int d^4l_1d^4l_2\hat{g}(l_1)\hat{g}(l_2)^{\ast}\int\frac{d^3\vec{k}}
               {2\om(\vec{k})}\frac{\abs{q_f}^2d\abs{q_f}d\Omega}{2\om(\vec{q_f})}\frac{1}{2\om(\vec{q_f}+\vec{k})}\\
             & \ \frac{F_1(\abs{q_f},\abs{k},\abs{k}\cos\theta)}{(q_f-p_i)^22q_f(l_1-k)(q_f-p_i)^22q_f(l_2-k)}\sum_{j=1}^{2}\frac{1}{\lvert f^{\prime}(\abs{q_f})
               \rvert_{\abs{q_f}=\abs{q_f}^{(j)}}\rvert}\delta(\abs{q_f}-\abs{q_f}^{(j)})\\
  \end{split}
\end{equation}
From this expression we can identify the differential cross section $\frac{d\sigma}{d\Omega}$. We do the $\abs{q_f}$-integration and obtain
\begin{equation}
  \begin{split}
    \frac{d\sigma}{d\Omega} = & \ \frac{1}{4}\frac{(2\pi)^{-5}\kappa^6}{\sqrt{\bigl((p_iq_i)^2-m^4\bigr)}}\int d^4l_1d^4l_2\hat{g}(l_1)\hat{g}(l_2)^{\ast}\int\frac{d^3\vec{k}}
                                {2\om(\vec{k})}\sum_{j=1}^{2}\frac{1}{\lvert f^{\prime}(\abs{q_f}^{(j)})\rvert}\\
                              & \ \frac{(\abs{q_f}^{(j)})^2}{2\sqrt{\bigl((\abs{q_f}^{(j)})^2+m^2\bigr)}}\frac{F_1(\abs{q_f}^{(j)},\abs{k},\abs{k}\cos\theta)}{2\sqrt{\bigl(
                                (\abs{q_f}^{(j)})^2+\abs{k}^2+\abs{k}\abs{q_f}^{(j)}\cos\theta+m^2\bigr)}}\\
                              & \ \Biggl(\frac{1}{m^2-2\sqrt{(\abs{q_f}^{(j)})^2+m^2}\om(\vec{p_i})+2\abs{q_f}^{(j)}\abs{p_i}\cos\al+p_i^2}\Biggr)^2\\
                              & \ \frac{1}{2\bigl[\sqrt{(\abs{q_f}^{(j)})^2+m^2}l_1^0-\abs{q_f}^{(j)}\abs{l_1}\cos\be-\sqrt{(\abs{q_f}^{(j)})^2+m^2}\abs{k}+\abs{q_f}^{(j)}
                                \abs{k}\cos\theta\bigr]}\\
                              &  \ \frac{1}{2\bigl[\sqrt{(\abs{q_f}^{(j)})^2+m^2}l_2^0-\abs{q_f}^{(j)}\abs{l_2}\cos\ga-\sqrt{(\abs{q_f}^{(j)})^2+m^2}\abs{k}+\abs{q_f}^{(j)}
                                \abs{k}\cos\theta\bigr]}\\
  \end{split}
\end{equation}
where we have introduced
\begin{equation}
  \begin{split}
    \al = & \ \sphericalangle(\vec{q_f},\vec{p_i})\\
    \be = & \ \sphericalangle(\vec{q_f},\vec{l_1})\\
    \ga = & \ \sphericalangle(\vec{q_f},\vec{l_2})\\
    \abs{l_i} = & \ |\vec{l_i}|,\ i=1,2\\
  \end{split}
\end{equation}
In the next step we omit all small quantities in the numerator, and we set $\abs{k}=0$ in the denominators if there occur no singularities. Then the expression can be simplified to 
\begin{equation}
  \begin{split}
    \frac{d\sigma}{d\Omega} = & \ \frac{1}{4}\frac{(2\pi)^{-5}\kappa^6}{\sqrt{\bigl((p_iq_i)^2-m^4\bigr)}}\frac{\om(\vec{p_i})}{2\abs{p_i}}\int d^4l_1d^4l_2\hat{g}(l_1)
                                \hat{g}(l_2)^{\ast}\int\frac{d^3\vec{k}}{2\om(\vec{k})}\\
                              & \ \sum_{j=1}^{2}\frac{\abs{p_i}^2}{4\om(\vec{p_i})^2}F_1(\abs{q_f}^{(j)}|_{\abs{k}=0},0,0)\Biggl(\frac{1}{m^2-2\om(\vec{p_i})^2
                                +2\abs{q_f}^{(j)}|_{\abs{k}=0}\abs{p_i}\cos\al+p_i^2}\Biggr)^2\\
                              & \ \frac{1}{2\bigl[\om(\vec{p_i})l_1^0-\abs{q_f}^{(j)}|_{\abs{k}=0}\abs{l_1}\cos\be-\om(\vec{p_i})\abs{k}+\abs{q_f}^{(j)}|_{\abs{k}=0}\abs{k}
                                \cos\theta\bigr]}\\
                              & \ \frac{1}{2\bigl[\om(\vec{p_i})l_2^0-\abs{q_f}^{(j)}|_{\abs{k}=0}\abs{l_2}\cos\ga-\om(\vec{p_i})\abs{k}+\abs{q_f}^{(j)}|_{\abs{k}=0}\abs{k}
                                \cos\theta\bigr]}\\
  \end{split}
\end{equation}
Now one observes that the function $F_1$ is independent of $\cos\theta$ and depends only on the initial momenta $p_i$ and $q_i$ so we can take it outside the integral. Then we get
\begin{equation}
  \begin{split}
    \frac{d\sigma}{d\Omega} = & \ \frac{1}{4}\frac{(2\pi)^{-5}\kappa^6}{\sqrt{\bigl((p_iq_i)^2-m^4\bigr)}}\sum_{j=1}^{2}F_2(\abs{q_f}^{(j)}|_{\abs{k}=0})\int d^4l_1d^4l_2
                                \hat{g}(l_1)\hat{g}(l_2)^{\ast}\\
                              & \ \int\frac{d^3\vec{k}}{2\om(\vec{k})}\frac{1}{2\bigl[\om(\vec{p_i})(l_1^0-\abs{k})-\abs{q_f}^{(j)}|_{\abs{k}=0}(\abs{l_1}\cos\be-\abs{k}
                                \cos\theta)\bigr]}\\
                              & \ \frac{1}{2\bigl[\om(\vec{p_i})(l_2^0-\abs{k})-\abs{q_f}^{(j)}|_{\abs{k}=0}(\abs{l_2}\cos\ga-\abs{k}
                                \cos\theta)\bigr]}\\
  \end{split}
\label{dsdO}
\end{equation}
where the function $F_2$ is given by
\begin{equation}
  F_2(\abs{q_f}^{(j)}|_{\abs{k}=0})=\frac{\abs{p_i}}{8\om(\vec{p_i})}F_1(\abs{q_f}^{(j)}|_{\abs{k}=0},0,0)\Biggl(\frac{1}{m^2-2\om(\vec{p_i})^2+2\abs{q_f}^{(j)}|_{\abs{k}=0}
                                    \abs{p_i}\cos\al+p_i^2}\Biggr)^2
\end{equation}
If we use (\ref{limitqf}) we can write the differential cross section in the form
\begin{equation}
  \begin{split}
    \frac{d\sigma}{d\Omega} = & \ \frac{1}{16}\frac{(2\pi)^{-5}\kappa^6}{\sqrt{\bigl((p_iq_i)^2-m^4\bigr)}}\int d^4l_1d^4l_2\hat{g}(l_1)\hat{g}(l_2)^{\ast}\Bigg[F_2(+\abs{p_i})
                                \int\frac{d^3\vec{k}}{2\om(\vec{k})}\\
                              & \ \frac{1}{\bigl[\om(\vec{p_i})(l_1^0-\abs{k})-\abs{p_i}(\abs{l_1}\cos\be-\abs{k}\cos\theta)\bigr]\bigl[\om(\vec{p_i})(l_2^0
                                -\abs{k})-\abs{p_i}(\abs{l_2}\cos\ga-\abs{k}\cos\theta)\bigr]}\\
                              & +F_2(-\abs{p_i})\int\frac{d^3\vec{k}}{2\om(\vec{k})}\\ 
                              & \ \frac{1}{\bigl[\om(\vec{p_i})(l_1^0-\abs{k})+\abs{p_i}(\abs{l_1}\cos\be-\abs{k}\cos\theta)\bigr]\bigl[\om(\vec{p_i})(l_2^0
                                -\abs{k})+\abs{p_i}(\abs{l_2}\cos\ga-\abs{k}\cos\theta)\bigr]}\Biggr]\\
  \end{split}
\label{diffcrosssec}
\end{equation}
In order to do the integration w.r.t. $\vec{k}$ we use an identity due to Feynman
\begin{equation}
  \frac{1}{ab}=\int_0^1\frac{dx}{[a x+b(1-x)]^2},\ a,b\in \mathbb{C}
\end{equation}
We consider the first term in (\ref{diffcrosssec}). Then we define the functions $a$ and $b$ by 
\begin{equation}
  \begin{split}
    a = & \ \bigl[\om(\vec{p_i})(l_1^0-\abs{k})-\abs{p_i}(\abs{l_1}\cos\be-\abs{k}\cos\theta)\bigr] \\
    b = & \ \bigl[\om(\vec{p_i})(l_2^0-\abs{k})-\abs{p_i}(\abs{l_2}\cos\ga-\abs{k}\cos\theta)\bigr] \\
  \end{split}
\label{ab}
\end{equation}
The integral becomes 
\begin{equation}
  \begin{split}
    \int\frac{d^3\vec{k}}{2\om(\vec{k})}\frac{1}{ab} = & \ \frac{1}{2}\int\abs{k}d\abs{k}d\Omega_{\abs{k}}\int_0^1\frac{dx}{\bigl[ax+b(1-x)\bigr]^2}\\
                                                     = & \ \frac{2\pi}{2}\int\abs{k}d\abs{k}\int_0^1dx\int_{-1}^{+1}\frac{d\cos\theta}{\bigl[ax+b(1-x)\bigr]^2}\\
  \end{split}
\end{equation}
We insert $a$ and $b$ and obtain
\begin{equation}
  \begin{split}
    = & \ \pi\int\abs{k}d\abs{k}\int_0^1dx\int_{-1}^{+1}\\
      & \ \frac{d\cos\theta}{\bigl[-\om(\vec{p_i})\abs{k}+\abs{p_i}\abs{k}\cos\theta+(\om(\vec{p_i})l_1^0-\abs{p_i}\abs{l_1}\cos\be)x+(\om(\vec{p_i})l_2^0-\abs{p_i}
        \abs{l_2}\cos\ga)(1-x)\bigr]^2}\\
  \end{split}
\label{theta-int}
\end{equation}
We set $\cos\theta=z$ and we observe that the integral w.r.t. $z$ is of the form 
\begin{equation}
  \int_{-1}^{+1}\frac{dz}{(cz+d)^2}=\frac{2}{d^2-c^2}
\label{bs1}
\end{equation}
see \cite{bs:tdm}, where we've set
\begin{equation}
  \begin{split}
    c = & \ \abs{p_i}\abs{k}\\
    d = & \ -\om(\vec{p_i})\abs{k}+\bigl(\om(\vec{p_i})l_1^0-\abs{p_i}\abs{l_1}\cos\be\bigr)x+\bigl(\om(\vec{p_i})l_2^0-\abs{p_i}\abs{l_2}\cos\ga\bigr)(1-x)\\
  \end{split}
\end{equation}
Then we obtain for (\ref{theta-int})
\begin{equation}
    = 2\pi\int\abs{k}d\abs{k}\int_0^1dx\frac{1}{p_i^2\abs{k}^2+A(l_1,l_2,x)\abs{k}+B(l_1,l_2,x)}    
\end{equation}
where $A(l_1,l_2,x)$ and $B(l_1,l_2,x)$ are given by
\begin{equation}
  \begin{split}
    A(l_1,l_2,x) = & -2\om(\vec{p_i})\Bigl[\bigl(\om(\vec{p_i})l_1^0-\abs{p_i}\abs{l_1}\cos\be\bigr)x+\bigl(\om(\vec{p_i})l_2^0-\abs{p_i}\abs{l_2}\cos\ga\bigr)(1-x)\Bigr]\\
    B(l_1,l_2,x) = & \ \Bigl[\bigl(\om(\vec{p_i})l_1^0-\abs{p_i}\abs{l_1}\cos\be\bigr)x+\bigl(\om(\vec{p_i})l_2^0-\abs{p_i}\abs{l_2}\cos\ga\bigr)(1-x)\Bigr]^2\\
  \end{split}
\end{equation}
Then this integral w.r.t. $\abs{k}$ is of the form
\begin{equation}
  \int \frac{ydy}{c_1y^2+c_2y+c_3}=\frac{1}{2c_1}\ln\bigl(c_1y^2+c_2y+c_3\bigr)-\frac{c_2}{2c_1}\int\frac{dy}{c_1y^2+c_2y+c_3}
\label{bs44}
\end{equation}
where the constants $c_i,\ i=1,2,3$ are given by
\begin{equation}
  \begin{split}
    c_1 = & \ p_i^2=m^2\\
    c_2 = & \ A(l_1,l_2,x)\\
    c_3 = & \ B(l_1,l_2,x)\\
  \end{split}
\end{equation}
In our case the discriminant $\Delta:=4c_1c_3-c_2^2$ is given by
\begin{equation}
  \begin{split}
    \Delta = & +4p_i^2\Bigl[\bigl(\om(\vec{p_i})l_1^0-\abs{p_i}\abs{l_1}\cos\be\bigr)x+\bigl(\om(\vec{p_i})l_2^0-\abs{p_i}\abs{l_2}\cos\ga\bigr)(1-x)\Bigr]^2\\
             & -4\om(\vec{p_i})^2\Bigl[\bigl(\om(\vec{p_i})l_1^0-\abs{p_i}\abs{l_1}\cos\be\bigr)x+\bigl(\om(\vec{p_i})l_2^0-\abs{p_i}\abs{l_2}\cos\ga\bigr)(1-x)\Bigr]^2\\
           = & \ -4\abs{p_i}^2\Bigl[\bigl(\om(\vec{p_i})l_1^0-\abs{p_i}\abs{l_1}\cos\be\bigr)x+\bigl(\om(\vec{p_i})l_2^0-\abs{p_i}\abs{l_2}\cos\ga\bigr)(1-x)\Bigr]^2\\
  \end{split}
\end{equation}
so we see that $\Delta\leq 0$. Then the integral (\ref{bs44}) is given by
\begin{equation}
  \int \frac{ydy}{c_1y^2+c_2y+c_3} =  \frac{1}{2c_1}\ln(c_1y^2+c_2y+c_3)-\frac{c_2}{2c_1\sqrt{-\Delta}}\ln\Biggl[\frac{2c_1y+c_2-\sqrt{-\Delta}}{2c_1y+c_2
                                      +\sqrt{-\Delta}}\Biggr]
\end{equation}
Here we consider gravitons up to an energy $\om_0$ which is much smaller than the energy of the incident particles. Since every real detector has a finite energy resolution below which he cannot detect any particles we have to integrate over all these contributions up to the value $\om_0$. Then the final result of the $\abs{k}$-integration is given by the following expression
\begin{equation}
  \begin{split}
    \int_0^{\om_0}\frac{\abs{k}d\abs{k}}{p_i^2\abs{k}^2+A\abs{k}+B} = & \ \frac{1}{2c_1}\Biggl[\ln\bigl(c_1\om_0^2+c_2\om_0+c_3\bigr)\\
                                                                      & -\frac{c_2}{\sqrt{-\Delta}}\ln\Biggl(\frac{2c_1\om_0+c_2-\sqrt{-\Delta}}{2c_1\om_0+c_2
                                                                        +\sqrt{-\Delta}}\Biggr)\\
                                                                      & -\ln(c_3)+\frac{c_2}{\sqrt{-\Delta}}\ln\Biggl(\Bigg\lvert\frac{c_2-\sqrt{-\Delta}}{c_2+\sqrt{-\Delta}}
                                                                        \Bigg\rvert\Biggr)\Biggr]\\
  \end{split}
\end{equation}
The second term in (\ref{diffcrosssec}) can be treated in the same way. The only difference is the sign of $\abs{p_i}$ in the definition of the functions $a$ and $b$, see (\ref{ab}), which has no consequences for the calculation of the integrals (\ref{bs1}) and (\ref{bs44}). 

%#*#*#*#*#*#*#*#*#*#*#*#*#*#*#*#*#*#*#*#*#*#*#*#*#*#*#*#*#*#*#*#*#*#*#*#*#*#*#*#*#*#*#*#*#*#*#*#*#*#*#*#*#*#*#*#*#*#*#*#*#*#*#*#*#*#*#*#*#*#*#*#*#*#*#*#*#
\subsection{Adiabatic limit}
In the preceeding subsection we've found the differential cross section for bremsstrahlung in a scattering process of two massive scalar particles. Now we want to discuss the adiabatic limit. For the differential cross section we've found
\begin{equation}
  \begin{split}
    \frac{d\sigma}{d\Omega} = & \ \frac{1}{16}\frac{(2\pi)^{-5}\kappa^6}{\sqrt{\bigl((p_iq_i)^2-m^4\bigr)}}\int d^4l_1d^4l_2\hat{g}(l_1)\hat{g}(l_2)^{\ast}
                                F_2(+\abs{p_i})\\
                              & \ 2\pi\int_0^1dx\frac{1}{2c_1}\Biggl[\ln\bigl(c_1\om_0^2+c_2\om_0+c_3\bigr)\\
                              & -\frac{c_2}{\sqrt{-\Delta}}\ln\Biggl(\frac{2c_1\om_0+c_2-\sqrt{-\Delta}}{2c_1\om_0+c_2+\sqrt{-\Delta}}\Biggr)\\
                              & \ -\ln(c_3)+\frac{c_2}{\sqrt{-\Delta}}\ln\Biggl(\Bigg\lvert\frac{c_2-\sqrt{-\Delta}}{c_2+\sqrt{-\Delta}}\Bigg\rvert\Biggr)\Biggr]\\
  \end{split}
\label{diffcrosssec1}
\end{equation}  
where $c_1=m^2$ and $c_i=c_i(l_1,l_2,x),\ i=2,3$. This expression, as it stands, is well defined and finite due to the presence of the testfunctions. To obtain a physically relevant result we have to remove this cutoff and therefore test the infrared behaviour of (\ref{diffcrosssec1}). We will show below that part of this expression becomes singular in the adiabatic limit. In the adiabatic limit in momentum space we let the testfunctions tend to a delta distribution in it's argument. This will be done in the same way as in subsection 5.1, see (\ref{adia}). We scale the argument of the testfunctions with a parameter $\vep$ in the integrals of (\ref{diffcrosssec1}) and consider the limit $\vep\rightarrow 0$. The functions $c_3$ and $\Delta$ are homogeneous of degree 2 in the variables $l_1$ and $l_2$, i.e. 
\begin{equation}
  \begin{split}
    c_3(\vep l_1,\vep l_2,x)=\vep^2 c_3(l_1,l_2,x)\\
    \Delta(\vep l_1,\vep l_2,x)=\vep^2\Delta(l_1,l_2,x)\\
  \end{split}
\end{equation}
while the function $c_2$ is homogeneous of degree 1, i.e. $c_2(\vep l_1,\vep l_2,x)=\vep c_2(l_1,l_2,x)$. With this in mind let us now look at the various logarithms in (\ref{diffcrosssec1}). The argument of the first logarithm contains the constant $m^2\om_0^2$ so it stays finite if $\vep$ goes to zero. The constant in front of the second and the fourth logarithm has equal powers of $\vep$ in the numerator as well as in the denominator so it tends to a constant. The argument of the second logarithm goes to one, so the logarithm itself goes to zero and the fourth goes to another constant. It remains to consider the third logarithm. This one becomes singular in the adiabatic limit. This can be seen as follows:
\begin{equation}
  \begin{split}
    \frac{d\sigma}{d\Omega} = & -\frac{\pi}{8}\frac{(2\pi)^{-5}\kappa^6}{\sqrt{\bigl((p_iq_i)^2-m^4\bigr)}}\int d^4l_1d^4l_2\hat{g}(l_1)\hat{g}(l_2)^{\ast}
                                F_2(+\abs{p_i})\frac{1}{2c_1}\int_0^1dx\\
                              & \ \ln\Biggl(\Bigl[\bigl(\om(\vec{p_i})l_1^0-\abs{p_i}\abs{l_1}\cos\be\bigr)x+\bigl(\om(\vec{p_i})l_2^0-\abs{p_i}\abs{l_2}\cos\ga\bigr)
                                (1-x)\Bigr]^2\Biggr)\\
                              & +\text{finite terms}\\
  \end{split}
\end{equation}
If we concentrate on the singular part only we obtain by inserting the scaled testfunctions 
\begin{equation}
  \begin{split}
    & \ \int d^4l_1d^4l_2\hat{g}(l_1)\hat{g}(l_2)^{\ast}\ln\bigl\lvert\bigl(\om(\vec{p_i})l_1^0-\abs{p_i}\abs{l_1}\cos\be\bigr)x+\bigl(\om(\vec{p_i})l_2^0
      -\abs{p_i}\abs{l_2}\cos\ga\bigr)(1-x)\bigr\rvert\\
  = & \ \int d^4l_1d^4l_2\hat{g}_{\vep}(l_1)\hat{g}_{\vep}(l_2)^{\ast}\ln\bigl\lvert\bigl(\om(\vec{p_i})l_1^0-\abs{p_i}\abs{l_1}\cos\be\bigr)x+\bigl(\om(\vec{p_i})l_2^0
      -\abs{p_i}\abs{l_2}\cos\ga\bigr)(1-x)\bigr\rvert\\
  = & \ \frac{1}{\vep^8}\int d^4l_1d^3l_2\hat{g}_0(\frac{l_1}{\vep})\hat{g}_0(\frac{l_2}{\vep})^{\ast}\ln\bigl\lvert\bigl(\om(\vec{p_i})l_1^0-\abs{p_i}\abs{l_1}\cos\be\bigr)x
      +\bigl(\om(\vec{p_i})l_2^0-\abs{p_i}\abs{l_2}\cos\ga\bigr)\\
    & \ \times(1-x)\bigr\rvert\\
  = & \int d^4k_1d^4k_2\hat{g}_0(k_1)\hat{g}_0(k_2)^{\ast}\ln\bigl\lvert\bigl(\om(\vec{p_i})\vep k_1^0-\abs{p_i}\vep\abs{k_1}\cos\be\bigr)x+\bigl(\om(\vec{p_i})\vep k_2^0-
      \abs{p_i}\vep\abs{k_2}\cos\ga\bigr)\\
    & \ \times(1-x)\bigr\rvert\\
  = & \ (2\pi)^4\ln\lvert\vep\rvert+o(1)\\
  \end{split}
\end{equation}
where we've introduced $k_i=\frac{l_i}{\vep}$. So we get a logarithmic divergence if $\vep$ tends to zero. The origin of this singularity is the massless graviton propagator as can be clearly seen from the above calculation. Since all the other terms in (\ref{T3}) have the same structure of propagators we would get the same logarihmic divergence as in our example. Furthermore there can be no cancellation between these terms since they all have a positive sign. So we can omit the discussion of the other terms. 

This result is similar to the case of quantum electrodynamics \cite{sch:qed} where a logarithmic divergence in the differential cross section was obtained as well in the bremsstrahlung contribution to the scattering of an electron due to an external classical source. As is well known from QED, the infrared divergence in the bremsstrahlung must cancel against contributions from radiative corrections, otherwise scattering theory would break down.  
  
\section*{Appendix}
In this appendix we give the definition of the singular order $\om$ of a numerical distribution $t\in\mathcal{S}(\mathbf{R}^4)$. The singular order measures the singularity of the distribution $t$ at the origin \cite{stei:axft}. Let $t\in\mathcal{S}(\mathbf{R}^4)$ be given. Then the distribution $t$ has scaling degree $s$ at $x=0$, if
\begin{equation*}
  s=\inf\{s^{\prime}\in\mathbb{R}|\lambda^{s^{\prime}}t(\lambda x)\stackrel{\lambda\searrow 0}{\longrightarrow}0\ \text{in the sense of distributions}\} 
\end{equation*}
The singular order is then defined by $\om:=[s]-4$, where $[s]$ is the greatest integer less or equal to $s$. It's an easy excercise to see that differentiation increases the singular order of a distribution, i.e.
\begin{equation*}
  \om(D^{a}t)=\om(t)+|a|
\end{equation*}
where $D^{a}=\partial^{a_1}_{x^0}\ldots\partial^{a_4}_{x^3}$ for every multi-index $a=(a_1,\ldots a_4)$ with $|a|=a_1+\ldots +a_4$.
  
%***************************************************************   L i t e r a t u r e   *****************************************************************************
%\bibliography{physics}
%\bibliographystyle{amsplain}
\input{ir1.bbl}

\end{document}

%% file: ir1.bbl
\providecommand{\bysame}{\leavevmode\hbox to3em{\hrulefill}\thinspace}